\documentclass[12pt]{article}
\usepackage{amsmath,float,amssymb,amsthm,amsxtra,overpic,bbm,bm,
epsfig,color}

\def\thefootnote{\fnsymbol{footnote}}

\usepackage{slashed,stmaryrd}
\usepackage{color}

\begin{document}

\vspace{0.2cm}

\begin{center}
{\Large\bf Looking into analytical approximations for three-flavor neutrino oscillation probabilities in matter}
\end{center}

\vspace{0.2cm}

\begin{center}
{\bf Yu-Feng Li $^{a,b}$}, \footnote{E-mail: liyufeng@ihep.ac.cn}
\quad
{\bf Jue Zhang $^{c}$}, \footnote{E-mail: juezhang87@pku.edu.cn}
\quad
{\bf Shun Zhou $^{a,b,c}$}, \footnote{E-mail: zhoush@ihep.ac.cn}
\quad
{\bf Jing-yu Zhu $^{a,b}$} \footnote{E-mail: zhujingyu@ihep.ac.cn}
\\
{\small $^a$Institute of High Energy Physics, Chinese Academy of
Sciences, Beijing 100049, China \\
$^b$School of Physical Sciences, University of Chinese Academy of Sciences, Beijing 100049, China \\
$^c$Center for High Energy Physics, Peking University, Beijing 100871, China}
\end{center}

\vspace{1.5cm}

\begin{abstract}
Motivated by tremendous progress in neutrino oscillation experiments, we derive a new set of simple and compact formulas for three-flavor neutrino oscillation probabilities in matter of a constant density. A useful definition of the $\eta$-gauge neutrino mass-squared difference $\Delta^{}_* \equiv \eta \Delta^{}_{31} + (1-\eta) \Delta^{}_{32}$ is introduced, where  $\Delta^{}_{ji} \equiv m^2_j - m^2_i$ for $ji = 21, 31, 32$ are the ordinary neutrino mass-squared differences and $0 \leq \eta \leq 1$ is a real and positive parameter. Expanding neutrino oscillation probabilities in terms of $\alpha \equiv \Delta^{}_{21}/\Delta^{}_*$, we demonstrate that the analytical formulas can be remarkably simplified for $\eta = \cos^2 \theta^{}_{12}$, with $\theta_{12}^{}$ being the solar mixing angle. As a by-product, the mapping from neutrino oscillation parameters in vacuum to their counterparts in matter is obtained at the order of ${\cal O}(\alpha^2)$. Finally, we show that our approximate formulas are not only valid for an arbitrary neutrino energy and any baseline length, but also still maintaining a high level of accuracy.
\end{abstract}

\begin{flushleft}
\hspace{0.8cm} PACS number(s): 14.60.Pq, 25.30.Pt
\end{flushleft}

\def\thefootnote{\arabic{footnote}}
\setcounter{footnote}{0}

\newpage

\section{Introduction}

Thanks to a number of elegant neutrino oscillation experiments in the past few decades, it has been well established that neutrinos are massive and lepton flavors are mixed~\cite{Agashe:2014kda}. In the framework of three-flavor neutrino oscillations, the lepton flavor mixing is described by a $3\times 3$ unitary matrix $U$, i.e., the Pontecorvo-Maki-Nakagawa-Sakata (PMNS) matrix~\cite{Pontecorvo:1957cp, Maki:1962mu}, which is conventionally parametrized in terms of three mixing angles $\{\theta^{}_{12}, \theta^{}_{13}, \theta^{}_{23}\}$ and one Dirac CP-violating phase $\delta$, namely,
\begin{eqnarray}
U = \begin{pmatrix} c^{}_{\theta^{}_{12}} c^{}_{\theta^{}_{13}} & s^{}_{\theta^{}_{12}} c^{}_{\theta^{}_{13}} & s^{}_{\theta^{}_{13}} e^{-{\rm i} \delta} \cr -s^{}_{\theta^{}_{12}} c^{}_{\theta^{}_{23}} - c^{}_{\theta^{}_{12}} s^{}_{\theta^{}_{13}} s^{}_{\theta^{}_{23}} e^{{\rm i} \delta} & c^{}_{\theta^{}_{12}} c^{}_{\theta^{}_{23}} - s^{}_{\theta^{}_{12}} s^{}_{\theta^{}_{13}} s^{}_{\theta^{}_{23}} e^{{\rm i} \delta} & c^{}_{\theta^{}_{13}} s^{}_{\theta^{}_{23}} \cr
s^{}_{\theta^{}_{12}} s^{}_{\theta^{}_{23}} - c^{}_{\theta^{}_{12}} s^{}_{\theta^{}_{13}} c^{}_{\theta^{}_{23}} e^{{\rm i}\delta} & ~ -c^{}_{\theta^{}_{12}} s^{}_{\theta^{}_{23}} - s^{}_{\theta^{}_{12}} s^{}_{\theta^{}_{13}} c^{}_{\theta^{}_{23}} e^{{\rm i} \delta} ~ & c^{}_{\theta^{}_{13}} c^{}_{\theta^{}_{23}} \cr
\end{pmatrix} \; ,
\label{eq:PMNS}
\end{eqnarray}
where $s^{}_{\theta^{}_{ij}} \equiv \sin \theta^{}_{ij}$ and $c^{}_{\theta^{}_{ij}} \equiv \cos \theta^{}_{ij}$ for $ij = 12, 13, 23$ have been defined.\footnote{If massive neutrinos are Majorana particles, two extra CP-violating phases are needed to parameterize the PMNS matrix, but they are irrelevant for neutrino oscillations.} The global-fit analysis of solar, atmospheric, reactor and accelerator neutrino oscillation experiments~\cite{Gonzalez-Garcia:2014bfa, Gonzalez-Garcia:2015qrr,Capozzi:2016rtj} yields three mixing angles $\theta^{}_{12} \approx 33^\circ$, $\theta^{}_{13} \approx 8.4^\circ$, $\theta^{}_{23} \approx 41^\circ$ and two neutrino mass-squared differences $\Delta^{}_{21} \equiv m^2_2 - m^2_1 \approx 7.4\times 10^{-5}~{\rm eV}^2$ and $\Delta^{}_{31} \equiv m^2_3 - m^2_1 \approx 2.5\times 10^{-3}~{\rm eV}^2$ in the case of normal mass ordering $m^{}_1 < m^{}_2 < m^{}_3$ (NMO), while $\theta^{}_{12} \approx 33^\circ$, $\theta^{}_{13} \approx 8.5^\circ$, $\theta^{}_{23} \approx 49^\circ$ and $\Delta^{}_{21} \approx 7.4\times 10^{-5}~{\rm eV}^2$, $\Delta^{}_{31} \approx -2.4\times 10^{-3}~{\rm eV}^2$ in the case of inverted mass ordering $m^{}_3 < m^{}_1 < m^{}_2$ (IMO). See Table 1 for a summary of the latest global-fit results from Ref.~\cite{Capozzi:2016rtj}.

Besides precision measurements of the known mixing parameters, the primary goals of ongoing and forthcoming oscillation experiments are to pin down neutrino mass ordering (i.e., the sign of $\Delta^{}_{31}$), to measure the leptonic CP-violating phase $\delta$, and to determine the octant of $\theta^{}_{23}$ (i.e., $\theta^{}_{23} < 45^\circ$ or $\theta^{}_{23} > 45^\circ$). In order to study the experimental sensitivities and better understand future experimental results, we should pay particular attention to the Mikheyev-Smirnov-Wolfenstein (MSW) matter effects on the propagation of neutrino beams in a medium~\cite{Wolfenstein:1977ue,Mikheev:1986gs}. Roughly speaking, current and future neutrino oscillation experiments can be categorized into three different types, in which terrestrial matter effects on neutrino oscillations always play an important role.
\begin{table}[t]
\centering
\begin{tabular}{ccccccccc}\hline \hline
&& \multicolumn{3}{c}{Normal mass ordering (NMO)} &&
\multicolumn{3}{c}{Inverted mass ordering (IMO)} \\ \hline
&& best-fit & & $3\sigma$ range & & best-fit & & $3\sigma$ range
\\ \hline
$\theta^{}_{12}$ && $33.02^\circ$ && $30^\circ$ --- $36.51^\circ$
&& $33.02^\circ$ && $30^\circ$ --- $36.51^\circ$ \\
$\theta^{}_{13}$ && $8.41^\circ$ && $7.82^\circ$ ---
$9.02^\circ$ && $8.49^\circ$ && $7.84^\circ$ --- $9.06^\circ$ \\
$\theta^{}_{23}$ && $41.38^\circ$ && $38^\circ$ ---
$51.71^\circ$ && $48.97^\circ$ && $38.23^\circ$ --- $52.95^\circ$ \\
$\delta$ && $243^\circ$ && $0^\circ$ --- $360^\circ$ && $237.6^\circ$
&& $0^\circ$ --- $360^\circ$ \\
\hline \\ \vspace{-1.15cm} \\
$\displaystyle \frac{\Delta^{}_{21}}{10^{-5} ~{\rm eV}^2}$ && $7.37$
&&$6.93$ --- $7.97$ && $7.37$ && $6.93$ --- $7.97$ \\ \vspace{-0.4cm} \\
$\displaystyle \frac{\Delta^{}_{31}}{10^{-3} ~{\rm eV}^2}$ &&
$2.537$ && $2.405$ --- $2.67$ && $-2.423$ && $-2.565$ --- $-2.29$ \\
\vspace{-0.47cm} \\ \hline \hline
\end{tabular}
\vspace{0.2cm}
\caption{The best-fit values and $3\sigma$ ranges of two neutrino mass-squared differences $\Delta^{}_{21}$ and $\Delta^{}_{31}$, three mixing angles $\{\theta^{}_{12}, \theta^{}_{13}, \theta^{}_{23}\}$ and the CP-violating phase $\delta$ from a global fit of current experimental data~\cite{Capozzi:2016rtj}.}
\end{table}
\begin{itemize}
\item {\it Medium-baseline Reactor Neutrino Experiments} --- The reactor experiments with a baseline length $L \approx 50~{\rm km}$ and a neutrino-beam energy $E \approx 4~{\rm MeV}$, such as JUNO~\cite{Li:2013zyd} and RENO-50~\cite{Kim:2014rfa}, are sensitive to the oscillations driven by both $\Delta^{}_{21}$ and $\Delta^{}_{31}$. Hence, they will be able to determine neutrino mass ordering and precisely measure oscillation parameters. It has been found~\cite{Li:2016txk} that the Earth matter effects for JUNO are as large as $1\%$, significantly affecting the determination of $\sin^2 \theta^{}_{12}$ and $\Delta^{}_{21}$, whose precisions are estimated to be $0.54\%$ and $0.24\%$, respectively~\cite{An:2015jdp}.

\item {\it Long-baseline Accelerator Neutrino Experiments} --- For the long-basline accelerator experiments T2K~\cite{T2K16} ($L = 295~{\rm km}$ and $E \approx 0.6~{\rm GeV}$), NO$\nu$A~\cite{NOvA16} ($L = 810~{\rm km}$ and $E \approx 2~{\rm GeV}$) and LBNF-DUNE~\cite{Acciarri:2015uup} ($L = 1300~{\rm km}$ and $E \approx 3~{\rm GeV}$), it is the relative sign between the matter potential for electron neutrinos (or antineutrinos) and $\Delta^{}_{31}$ that changes the oscillation probability of $\nu^{}_\mu \to \nu^{}_e$ (or $\overline{\nu}^{}_\mu \to \overline{\nu}^{}_e$), opening another possibility to pin down neutrino mass ordering. The difference between oscillation probabilities of neutrinos and those of antineutrinos implies leptonic CP violation, which however suffers from a contamination induced by the CP-asymmetric Earth matter. In addition, the neutrino super-beam experiments ESS$\nu$SB ($L \approx 500~{\rm km}$ and $0.2~{\rm GeV} \lesssim E \lesssim 0.6~{\rm GeV}$) and MOMENT ($L \approx 150~{\rm km}$ and $0.15~{\rm GeV} \lesssim E \lesssim 0.20~{\rm GeV}$) have also been proposed to measure the CP-violating phase with relatively low energy neutrinos and short baseline lengths~\cite{Baussan:2013zcy,Cao:2014bea,Blennow:2015cmn} .

\item {\it Huge Atmospheric Neutrino Experiments} --- The experiments PINGU~\cite{Aartsen:2014oha}, ORCA~\cite{Katz:2014tta}, and Hyper-Kamiokande~\cite{Abe:2011ts} will implement huge ice or water Cherenkov detectors to precisely measure atmospheric neutrinos,
    for which a wide range of energies ($0.1~{\rm GeV} \lesssim E \lesssim 100~{\rm GeV}$) and baseline lengths ($10~{\rm km} \lesssim L \lesssim 10^4~{\rm km}$) should be considered. Though neutrinos and antineutrinos cannot be distinguished in these experiments, the MSW resonance in the Earth matter occurs either in neutrino oscillations for NMO or in antineutrino oscillations for IMO. Therefore, matter effects help determine neutrino mass ordering. A $3\sigma$ significance can be reached at the ICAL detector of INO, which can also discriminate between $\nu^{}_\mu$ and $\overline{\nu}^{}_\mu$ events~\cite{Ahmed:2015jtv}.
\end{itemize}
In principle, for any neutrino energy and baseline length, one can exactly calculate neutrino and antineutrino oscillation probabilities in the Earth matter by numerical methods. However, it is obviously difficult in this way to reveal the underlying physics for neutrino oscillations and to fully understand the numerical results.

For this reason, two theoretical approaches have been suggested to study the Earth matter effects. First, one can establish an exact relation between the effective parameters in matter, i.e., the mixing matrix $\widetilde{U}$ (parametrized in terms of three mixing angles $\widetilde{\theta}^{}_{ij}$ and one CP-violating phase $\widetilde{\delta}$) and three neutrino masses $\widetilde{m}^{}_i$ (or mass-squared differences $\widetilde{\Delta}^{}_{ji} \equiv \widetilde{m}^2_j - \widetilde{m}^2_i$), and the intrinsic parameters in vacuum. For example, $\widetilde{\cal J} \widetilde{\Delta}^{}_{21} \widetilde{\Delta}^{}_{31} \widetilde{\Delta}^{}_{32} = {\cal J} \Delta^{}_{21} \Delta^{}_{31} \Delta^{}_{32}$ holds exactly for a constant matter density~\cite{Naumov:1991ju,Harrison:1999df,Xing:2001bg}, where the Jarlskog invariant in vacuum~\cite{Jarlskog:1985ht,Wu:1985ea} is defined by ${\cal J} \sum_{\gamma, k} \epsilon^{}_{\alpha \beta \gamma} \epsilon^{}_{ijk} \equiv {\rm Im}\left[U^{}_{\alpha i} U^*_{\alpha j} U^*_{\beta i} U^{}_{\beta j}\right]$ with the Greek and Latin letters running over $(e, \mu, \tau)$ and $(1, 2, 3)$, respectively, and likewise for $\widetilde{\cal J}$ and $\widetilde{U}$. Another example is the Toshev relation $\sin 2\widetilde{\theta}^{}_{23} \sin \widetilde{\delta} = \sin 2\theta^{}_{23} \sin \delta$~\cite{Toshev:1991ku} in the standard parametrization. In addition, the notion of unitarity triangles has also been introduced to describe leptonic CP violation~\cite{Fritzsch:1999ee,AguilarSaavedra:2000vr,Sato:2000wv,Farzan:2002ct}, and the exact and approximate relations between the unitarity triangles in matter and those in vacuum have been found in Refs.~\cite{Zhang:2004hf,Xing:2005gk,He:2013rba,Xing:2015wzz,He:2016dco}. Although these exact relations are interesting in themselves, they are in practice not useful to directly explain experimental observations and extract fundamental oscillation parameters.

The second approach is to expand the oscillation probabilities in terms of small perturbation parameters, which can be $\alpha \equiv \Delta^{}_{21}/\Delta^{}_* \approx 0.03$ [where $\Delta^{}_* \equiv \eta \Delta^{}_{31} + (1-\eta) \Delta^{}_{32}$ with $0 \leq \eta \leq 1$, cf. Eq.~(\ref{eq:Dels})] and the smallest mixing angle $\sin\theta^{}_{13} \approx 0.147$~\cite{Cervera:2000kp,Freund:2001pn,Akhmedov:2004ny, Xu:2015kma,Minakata:2015gra,Denton:2016wmg,Agarwalla:2013tza,Flores:2015mah}. Another choice is $\widehat{A} \equiv A/\Delta^{}_*$, where $A$ is the matter potential from the coherent forward scattering of neutrinos on the background particles and defined as $A \equiv 2\sqrt{2}G^{}_{\rm F} N^{}_e E$, with $G^{}_{\rm F}$ being the Fermi constant, $N^{}_e$ the number density of background electrons, and $E$ the neutrino energy. In the case of low neutrino energies or low matter densities, an expansion in $\widehat{A}$ is useful to show the corrections of matter potential to the oscillation probabilities in vacuum.

In the seminal paper by Freund~\cite{Freund:2001pn}, the analytical approximations for three-flavor neutrino oscillation probabilities have been systematically studied, and the formulas are valid as long as the oscillation driven by $\Delta^{}_{21}$ has not developed and the corresponding MSW resonance is not reached. The latter condition corresponds to $\widehat{A} \gtrsim \alpha$~\cite{Freund:2001pn}, namely,
\begin{eqnarray}
E \gtrsim 0.34~{\rm GeV}~\left(\frac{\Delta^{}_{21}}{7.5\times 10^{-5}~{\rm eV}^2}\right)\cdot \left(\frac{2.8~{\rm g}~{\rm cm}^{-3}}{\rho}\right) \; ,
\label{eq:cond}
\end{eqnarray}
where $\rho$ is the matter density. For the Earth matter, the electron fraction is $Y^{}_e \approx 0.5$ and $N^{}_e = Y^{}_e N^{}_{\rm A} [\rho/(1~{\rm g}~{\rm cm}^{-3})]$ with $N^{}_{\rm A}$ being the Avogadro's number.  Although Freund's formulas actually work even for $E < 0.34~{\rm GeV}$, it has been shown in Ref.~\cite{Xu:2015kma} and Ref.~\cite{Xing:2016ymg} that the series expansion of $\widehat{\epsilon} \equiv (\alpha^2 + \widehat{A}^2 \cos^4\theta^{}_{13} - 2\widehat{A}\alpha \cos2\theta^{}_{12} \cos^2\theta^{}_{13})^{1/2}$ in terms of $\alpha$ is problematic in the region of low energies or small matter densities, where $\widehat{A} \to 0$. More accurate approximate formulas for low energies $E < 1~{\rm GeV}$ have been derived in Ref.~\cite{Xing:2016ymg} by retaining $\widehat{\epsilon}$. However, the analytical results in Refs.~\cite{Xu:2015kma,Xing:2016ymg} are not applicable for large matter effects and higher neutrino energies. Furthermore, a critical problem for the $\sin\theta_{13}$ expansion is related to the atmospheric resonance $\widehat{A} \to 1$, where the function $\widehat{C} \equiv [(1-\widehat{A})^2 + 4\widehat{A} \sin^2\theta^{}_{13}]^{1/2}$ cannot be expanded correctly. As we will show later, $\widehat{\epsilon}$ and $\widehat{C}$ are two key parameters to avoid any difficulties associated with the low-energy solar resonance and the high-energy atmospheric resonance, respectively. In fact, analytical formulas for arbitrary neutrino energies and baseline lengths are derived in Refs.~\cite{Minakata:2015gra,Denton:2016wmg}, where the resonances related to $\Delta^{}_{21}$ and $\Delta^{}_{31}$ have been treated carefully by introducing a few intermediate rotation angles for basis transformations. Thus, the analytical results can be cast into a simple and compact form, in which the eigenvalues of the zeroth-order Hamiltonian and rotation angles, instead of intrinsic mixing parameters, are involved.

Since all the existing analytical approximations are not fully satisfactory, we are well motivated to derive a new set of analytical formulas for neutrino oscillation probabilities, which fulfills the following three criteria:
\begin{enumerate}
\item They are valid for arbitrary neutrino energies and any baseline length. Such formulas are applicable to atmospheric neutrino experiments.

\item They are expressed in terms of intrinsic oscillation parameters, and in a simple and compact form. Any complicated formulas are not very useful in practice.

\item They give accurate values of oscillation probabilities, under the condition that the first two criteria are met at the same time.
\end{enumerate}
For this purpose, we expand the oscillation probabilities in terms of $\alpha$, but retain the parameter that corresponds to $\widehat{\epsilon}$ ($\widehat{C}$) in the case of low (high) energies or small (large) matter densities. In addition, an $\eta$-gauge neutrino mass-squared difference $\Delta^{}_* \equiv \eta \Delta^{}_{31} + (1-\eta)\Delta^{}_{32}$ is introduced so as to seek an optimal value of $\eta$ that greatly simplifies approximate formulas.

The remaining part of this paper is organized as follows. In Section 2, we briefly review the basic strategy to derive analytical formulas of neutrino oscillation probabilities. We introduce $\Delta^{}_* \equiv \eta \Delta^{}_{31} + (1-\eta)\Delta^{}_{32}$ and demonstrate that it is suggestive of simple analytical formulas for $\eta = \cos^2 \theta^{}_{12}$. The oscillation probabilities in the special case of $\eta = \cos^2 \theta^{}_{12}$ are presented in Section 3, where the mapping between effective and intrinsic mixing parameters is also obtained as a by-product. The accuracies of the analytical formulas are examined and compared with previous ones. Finally, we summarize our main results in Section 4. Some useful formulas are listed in three appendices.

\section{General Formalism}

In the framework of three-flavor neutrino oscillations, the effective Hamiltonian responsible for the evolution of neutrino flavor eigenstates in matter is given by
\begin{eqnarray}
\widetilde{H}^{}_{\rm f} = \frac{1}{2 E} \left[U
\begin{pmatrix} m^2_1 & 0 & 0 \cr 0 & m^2_2 & 0 \cr 0 & 0 & m^2_3
\cr \end{pmatrix} U^\dagger + \begin{pmatrix} A & 0 & 0 \cr 0 & 0 &
0 \cr 0 & 0 & 0 \cr \end{pmatrix} \right] \; .
\label{eq:Heff_f}
\end{eqnarray}
In the case of a constant matter density, i.e., a constant value of $A$, we then have two distinct ways to derive the exact oscillation probabilities. First, one can diagonalize the effective Hamiltonian by using a unitary transformation
\begin{eqnarray}
\widetilde{H}^{}_{\rm f} = \frac{1}{2E} \widetilde{U} \begin{pmatrix} \widetilde{m}^2_1 & 0 & 0 \cr 0 & \widetilde{m}^2_2 & 0 \cr 0 & 0 & \widetilde{m}^2_3
\cr \end{pmatrix} \widetilde{U}^\dagger \; ,
\label{eq:Heff_diag}
\end{eqnarray}
where $\widetilde{m}^{}_i$ for $i = 1, 2, 3$ are effective neutrino masses in matter, and $\widetilde{U}$ is the effective PMNS matrix, which can also be parametrized in terms of three mixing angles $\{\widetilde{\theta}^{}_{12}, \widetilde{\theta}^{}_{13}, \widetilde{\theta}^{}_{23}\}$ and one CP-violating phase $\widetilde{\delta}$. In terms of these effective parameters, it is straightforward to write down the oscillation probabilities $\widetilde{P}^{}_{\alpha \beta} \equiv \widetilde{P}(\nu^{}_\alpha \to \nu^{}_\beta)$ as follows
\begin{eqnarray}
\widetilde{P}^{}_{\alpha \beta} = \delta^{}_{\alpha \beta} - 4 \sum^3_{i<j} {\rm Re}\left[\widetilde{U}^{}_{\alpha i} \widetilde{U}^*_{\alpha j} \widetilde{U}^*_{\beta i} \widetilde{U}^{}_{\beta j}\right] \sin^2 \widetilde{F}^{}_{ji}+ 8\widetilde{J} \sum_\gamma \epsilon^{}_{\alpha \beta \gamma} \sin \widetilde{F}^{}_{21} \sin \widetilde{F}^{}_{31} \sin \widetilde{F}^{}_{32} \; ,
\label{eq:Pab_eff}
\end{eqnarray}
where $\widetilde{J} \equiv \sum_{\gamma, k}\epsilon^{}_{\alpha \beta \gamma} \epsilon^{}_{ijk} {\rm Im}\left[\widetilde{U}^{}_{\alpha i } \widetilde{U}^*_{\alpha j} \widetilde{U}^*_{\beta i} \widetilde{U}^{}_{\beta j}\right]$ and $\widetilde{F}^{}_{ji} \equiv \widetilde{\Delta}^{}_{ji}L/(4E)$ with $\widetilde{\Delta}^{}_{ji} \equiv \widetilde{m}^2_j - \widetilde{m}^2_i$ have been defined in the same manner as for neutrino oscillations in vacuum, and $L$ is the baseline length. The probabilities for antineutrino oscillations $\overline{\nu}^{}_\alpha \to \overline{\nu}^{}_\beta$ can be obtained by replacing $\widetilde{J} \to -\widetilde{J}$ in Eq.~(\ref{eq:Pab_eff}) and $A \to -A$ everywhere in the effective parameters.

Second, according to the Cayley-Hamilton theorem, the evolution matrix $S = e^{-{\rm i} \widetilde{H}^{}_{\rm f} L}$ of neutrino flavor eigenstates is determined by three eigenvalues of the effective Hamiltonian $\widetilde{H}^{}_{\rm f}$ and the matrix elements of $\widetilde{H}^{}_{\rm f}$~\cite{Moler,Ohlsson:1999xb,Ohlsson:1999um}, namely,
\begin{eqnarray}
S^{}_{\beta \alpha} = s^{}_0 I^{}_{\beta \alpha} + s^{}_1 \left(\widetilde{H}^{}_{\rm f}\right)^{}_{\beta \alpha} + s^{}_2 \left(\widetilde{H}^2_{\rm f}\right)^{}_{\beta \alpha} \; ,
\label{eq:CH}
\end{eqnarray}
where $I$ denotes the $3\times 3$ unit matrix and the relevant coefficients are
\begin{eqnarray}
s_0^{} & = & - \frac{\omega^{}_{1} \omega^{}_{2}e^{-{\rm i}\omega_{3}^{} L} }{(\omega^{}_{2} - \omega^{}_{3})(\omega^{}_{3} - \omega^{}_{1})} - \frac{\omega^{}_{2} \omega^{}_{3}e^{-{\rm i} \omega_{1}^{} L} }{(\omega^{}_{1} - \omega^{}_{2})(\omega^{}_{3} - \omega^{}_{1})} - \frac{\omega^{}_{1} \omega^{}_{3}e^{-{\rm i} \omega_{2}^{} L} }{(\omega^{}_{1} - \omega^{}_{2})(\omega^{}_{2} - \omega^{}_{3})} \; , \nonumber \\
s_1^{} & = & + \frac{(\omega^{}_{1} + \omega^{}_{2}) e^{-{\rm i}\omega_{3}^{} L} }{(\omega^{}_{2} - \omega^{}_{3})(\omega^{}_{3} - \omega^{}_{1})} + \frac{(\omega^{}_{2} + \omega^{}_{3}) e^{-{\rm i} \omega_{1}^{} L} }{(\omega^{}_{1} - \omega^{}_{2})(\omega^{}_{3} - \omega^{}_{1})} + \frac{(\omega^{}_{1} + \omega^{}_{3}) e^{-{\rm i} \omega_{2}^{} L} }{(\omega^{}_{1} - \omega^{}_{2})(\omega^{}_{2} - \omega^{}_{3})} \; , \nonumber \\
s_2^{} & = & - \frac{e^{-{\rm i}\omega_{3}^{} L} }{(\omega^{}_{2} - \omega^{}_{3})(\omega^{}_{3} - \omega^{}_{1})} - \frac{e^{-{\rm i} \omega_{1}^{} L} }{(\omega^{}_{1} - \omega^{}_{2})(\omega^{}_{3} - \omega^{}_{1})} - \frac{e^{-{\rm i} \omega_{2}^{} L} }{(\omega^{}_{1} - \omega^{}_{2})(\omega^{}_{2} - \omega^{}_{3})} \; ,
\label{eq:si}
\end{eqnarray}
with $\omega^{}_i \equiv \widetilde{m}^2_i/(2E)$ being the eigenvalues of $\widetilde{H}^{}_{\rm f}$. The neutrino oscillation probabilities are simply given by $\widetilde{P}^{}_{\alpha \beta} = |S^{}_{\beta \alpha}|^2$, while the results for antineutrino oscillations can be derived by changing $U \to U^*$ and $A \to -A$ in the effective Hamiltonian.

\subsection{$\eta$-gauge Mass-squared Difference}

In order to simplify the analytical formulas as much as possible, we tentatively introduce a generic definition of neutrino mass-squared difference
\begin{eqnarray}
\Delta^{}_* \equiv \eta \Delta^{}_{31} + (1-\eta) \Delta^{}_{32}
\; ,
\label{eq:Dels}
\end{eqnarray}
where $0 \leq \eta \leq 1$ is a real and positive parameter. It is evident that $\Delta^{}_*$ reduces to the conventional definitions of atmospheric neutrino mass-squared differnce $\Delta^{}_{32}$ for $\eta = 0$, $\Delta^{}_{31}$ for $\eta = 1$, and $(\Delta^{}_{31}+\Delta^{}_{32})/2$ for $\eta = 1/2$. In the global-fit analysis of neutrino oscillation data, the first two definitions have been used in Ref.~\cite{Gonzalez-Garcia:2014bfa,Gonzalez-Garcia:2015qrr} in the IMO and NMO cases, respectively, while the last one has been implemented in Ref.~\cite{Capozzi:2016rtj} for either neutrino mass ordering. Another definition $\Delta^{}_{ee} \equiv \cos^2\theta^{}_{12} \Delta^{}_{31} + \sin^2\theta^{}_{12} \Delta^{}_{32}$, corresponding to $\eta = \cos^2\theta^{}_{12}$, has been advocated by Parke~\cite{Parke:2016joa} and demonstrated to be advantageous to reactor antineutrino experiments. Although for quite a different reason, as we will show later, the introduction of $\Delta^{}_*$ in Eq.~(\ref{eq:Dels}) with $\eta = \cos^2\theta^{}_{12}$ turns out to be very useful in simplifying the approximate formulas of oscillation probabilities.

With the help of $\Delta^{}_*$, the effective Hamiltonian $\widetilde{H}^{}_{\rm f}$ can be rewritten as
\begin{eqnarray}
\widetilde{H}^{}_{\rm f} = \frac{m_2^2-\eta \Delta^{}_{21}}{2E}I +
\frac{\Delta_{*}^{}}{2 E} M^{}_{\rm f}
\label{eq:Heff_new}
\; ,
\end{eqnarray}
where $I$ is the identity matrix of rank three and
\begin{eqnarray}
M^{}_{\rm f}   & = &
U \begin{pmatrix}
\left(\eta - 1\right) \alpha& 0 & 0 \cr
0 & \eta\alpha &  0 \cr  0&0&1\cr
\end{pmatrix} U^{\dag}_{} +
\begin{pmatrix}
\widehat{A} &0 & 0 \cr 0 &0 & 0 \cr 0 &0 & 0 \cr
\end{pmatrix} \; ,
\label{eq:M_f}
\end{eqnarray}
with $\alpha \equiv {\Delta_{21}^{}}/{\Delta_*^{}}$ and $\widehat{A} = {A}/{\Delta_*^{}}$. Note that the first term on the right-hand side of Eq.~(\ref{eq:Heff_new}) is flavor-independent and thus irrelevant for neutrino oscillations. In the formalism shown in Eqs.~(\ref{eq:CH}) and (\ref{eq:si}), the evolution matrix is now $S = e^{-{\rm i}\Delta^{}_* M^{}_{\rm f}L/(2E)}$ and only the eigenvalues of $M^{}_{\rm f}$ need to be calculated.

To find out the eigenvalues of $M^{}_{\rm f}$, it is more convenient to convert into the mass basis in vacuum via $M^{}_{\rm v} = U^\dagger M^{}_{\rm f} U$, where the neutrino mass term, i.e., the first term on the right-hand side of Eq.~(\ref{eq:M_f}), becomes diagonal. More explicitly, we have~\cite{Xing2000}
\begin{eqnarray}
M^{}_{\rm v} & = &
\begin{pmatrix}
\left(\eta - 1\right) \alpha& 0 & 0 \cr
0 & \eta \alpha &  0 \cr  0 & 0 & 1 \cr
\end{pmatrix}  +
\widehat{A} \begin{pmatrix}
|U^{}_{e1}|^2 & U^*_{e1} U^{}_{e2} & U^*_{e1} U^{}_{e3} \cr U^*_{e2} U^{}_{e1} & |U^{}_{e2}|^2 & U^*_{e2} U^{}_{e3} \cr U^*_{e3} U^{}_{e1} & U^*_{e3} U^{}_{e2} & |U^{}_{e3}|^2 \cr
\end{pmatrix} \; ,
\label{eq:M_v}
\end{eqnarray}
where $U^{}_{ei}$ for $i = 1, 2, 3$ are three elements in the first row of the PMNS matrix. Therefore, it is expected that a proper choice of $\eta$ will be helpful in reducing the complexity of three eigenvalues, and thus the final oscillation probabilities. Furthermore, it is interesting to notice that only the matrix elements $U^{}_{ei}$ for $i = 1, 2, 3$ are involved in $M^{}_{\rm v}$ and $|U^{}_{e3}| \ll 1$, so a suitable value of $\eta$ is anticipated to be mainly associated with $U^{}_{e1}$ and $U^{}_{e2}$, or $\theta^{}_{12}$ in the standard parametrization.

\subsection{$\eta$-gauge Oscillation Probabilities}

Now it is time to derive the oscillation probabilities by using Eqs.~(\ref{eq:CH}) and (\ref{eq:si}). First of all, the eigenvalues $\lambda_{i}$ (for $i=1,2,3$) of $M^{}_{\rm f}$ or equivalently $M^{}_{\rm v}$ can be obtained by solving the following eigen-equation
\begin{eqnarray}
\lambda_{}^{3} + b \lambda^2_{} + c \lambda + d = 0 \, ,
\label{eq:eigen}
\end{eqnarray}
where the relevant coefficients are
\begin{eqnarray}
b & = &
- 1 -\alpha \left( 2\eta -1 \right) -\widehat{A} \; ,\nonumber \\
c & = &
\left(1 - |U_{e3}^{}|^2\right) \widehat{A} - \alpha \left\{
1 + \widehat{A} |U_{e2}^{}|^2 + \widehat{A} |U_{e3}^{}|^2
- \eta\left[ 2 + \alpha \left(\eta - 1\right)
+ \widehat{A} + \widehat{A} |U_{e3}^{}|^2 \right]  \right\}
\; , \nonumber  \\
d & = &
- \alpha \left[ \widehat{A} \eta|U_{e1}^{}|^2  + \widehat{A}
\left( \eta - 1 \right) |U_{e2}^{}|^2  + \alpha \eta (\eta - 1)
\big( 1 + \widehat{A} |U_{e3}^{}|^2 \big)  \right] \; .
\label{eq:bcd}
\end{eqnarray}
The eigenvalues of the effective Hamiltonian have been known for a long time~\cite{Barger,Zaglauer,Xing2000}, but it has recently been noticed that the results depend also on neutrino mass ordering~\cite{Xing:2016ymg}. To be explicit, taking $\lambda^{}_1 < \lambda^{}_2 < \lambda^{}_3$, we have
\begin{eqnarray}
\lambda_{1}^{} &=& -\frac{b}{3} - \frac{1}{3\Delta_*^{}} \sqrt{x^2 - 3 y} \Big[ z + \sqrt{3(1-z^2_{})} \Big] \; , \nonumber\\
\lambda_{2}^{} &=& -\frac{b}{3} - \frac{1}{3\Delta_*^{}} \sqrt{x^2 - 3 y} \Big[ z - \sqrt{3(1-z^2_{})} \Big] \; , \nonumber\\
\lambda_{3}^{} &=& -\frac{b}{3} + \frac{2}{3\Delta_*^{}} z\sqrt{x^2 - 3y} \; ,
\label{eq:lambda_NMO}
\end{eqnarray}
for the NMO; or
\begin{eqnarray}
\lambda_{1}^{} &=& -\frac{b}{3} + \frac{1}{3\Delta_*^{}} \sqrt{x^2 - 3 y} \Big[ z + \sqrt{3(1-z^2_{})} \Big] \;, \nonumber\\
\lambda_{2}^{} &=& -\frac{b}{3} + \frac{1}{3\Delta_*^{}} \sqrt{x^2 - 3 y} \Big[ z - \sqrt{3(1-z^2_{})} \Big] \;, \nonumber\\
\lambda_{3}^{} &=& -\frac{b}{3} - \frac{2}{3\Delta_*^{}} z\sqrt{x^2 - 3y} \label{eq:lambda_IMO}
\; ,
\end{eqnarray}
for the IMO, where we have defined
\begin{eqnarray}
x &=& \Delta_{*}^{}\Big[1 + (2-\eta)\alpha
+ \widehat{A}  \Big] \;, \nonumber\\
y &=& \Delta_{*}^{2} \bigg\{\widehat{A}(1 - |U_{e3}^{}|^2) + \alpha  \Big[1 + \widehat{A} - \widehat{A} (1 - |U_{e3}^{}|^2)(\eta - \frac{|U_{e1}^{}|^2}{1 - |U_{e3}^{}|^2})\Big] + \alpha^2 (1 - \eta)
\bigg\} \;, \nonumber\\
z &=& \cos \bigg\{
\frac13 \arccos \frac{\Delta_*^{} \Big[2 x^3 - 9xy + 27\Delta_*^{3}
\alpha \widehat{A} (1+ \alpha - \eta \alpha) |U_{e1}^{}|^2\Big]}
{2 |\Delta_*^{}|\sqrt{(x^2 -3y)^3}}\bigg\} \; .
\label{eq:xyz}
\end{eqnarray}
Note that $\lambda^{}_i$'s are the eigenvalues of $M^{}_{\rm f}$, and $(\lambda^{}_1, \lambda^{}_2, \lambda^{}_3)$ correspond to $(\widetilde{m}^2_1, \widetilde{m}^2_2, \widetilde{m}^2_3)$ in the NMO case with $\Delta^{}_* > 0$, but to $(\widetilde{m}^2_2, \widetilde{m}^2_1, \widetilde{m}^2_3)$ in the IMO case with $\Delta^{}_* < 0$. In the latter case, though $\lambda^{}_3$ is the largest eigenvalue, $\Delta^{}_* \lambda^{}_3$ becomes negative and thus $\widetilde{m}^2_3$ is the smallest one. In addition, it is easy to verify that $\lambda^{}_2 - \lambda^{}_1 > 0$ holds for either neutrino mass ordering.

According to Eqs.~(\ref{eq:CH}) and (\ref{eq:si}), it is straightforward to compute the evolution matrix $S = e^{-2{\rm i}F^{}_* M^{}_{\rm f}}$ with $F^{}_* \equiv \Delta^{}_* L/(4E)$ and thus the oscillation probabilities
\begin{eqnarray}
\widetilde P_{ \alpha\beta}^{} & = & \left|\xi_1^{\alpha\beta} e^{- {\rm i}F_*^{}(2\lambda_{3}^{} - \lambda^{}_1 - \lambda^{}_2)} + \xi_2^{ \alpha\beta} \cos\big[F_*^{}(\lambda_{2}^{} - \lambda_{1}^{})\big] + 2 {\rm i} \xi_3^{ \alpha\beta} \frac{\sin\big[F_*^{}(\lambda_{2}^{} -
\lambda_{1}^{})\big]}{\lambda_{2}^{} - \lambda_{1}^{}}
\right|^2 \; ,
\label{eq:Pab_s}
\end{eqnarray}
where the flavor-dependent coefficients $\xi^{\alpha \beta}_i$ (for $i = 1, 2, 3$) with $\alpha$ and $\beta$ running over $e$, $\mu$ and $\tau$ can readily be identified from similar equations for $M^{}_{\rm f}$ to those for $\widetilde{H}^{}_{\rm f}$ in Eqs.~(\ref{eq:CH}) and (\ref{eq:si}). A further exploration of the right-hand side of Eq.~(\ref{eq:Pab_s}) gives rise to
\begin{eqnarray}
\widetilde P_{\alpha \beta}^{} &=& |\xi_1^{\alpha \beta}|^2 + |\xi_2^{\alpha \beta}|^2 +  \left\{
4 |\xi_3^{\alpha \beta}|^2 - (\lambda_{2}^{}
-\lambda_{1}^{})
^{2} |\xi_2^{\alpha \beta}|^2 \right\}
\frac{\sin^2\big[F_*^{}(\lambda_{2}^{} -
\lambda_{1}^{})\big]}{(\lambda_{2}^{} - \lambda_{1}^{})^2}
\nonumber \\
&~& +  2\left\{{\rm Re}[\xi^{\alpha \beta}_1 \xi^{\alpha \beta *}_2]
\cos\big[F_*^{} ( 3 \lambda_3^{} + b  ) \big] + {\rm Im}[\xi^{\alpha \beta }_1 \xi^{\alpha \beta*}_2] \sin\big[F_*^{}( 3 \lambda_3^{} + b  )    \big]
\right\}\cos \big[F_*^{}(\lambda_{2}^{} -
\lambda_{1}^{})\big] \nonumber \\
&~& + 4\left\{ {\rm Im}[\xi^{\alpha \beta}_1 \xi^{\alpha \beta *}_3]
\cos\big[F_*^{}( 3 \lambda_3^{} + b  )  \big] -{\rm Re}[\xi^{\alpha \beta }_1 \xi^{\alpha \beta*}_3]
\sin\big[F_*^{} ( 3 \lambda_3^{} + b  ) \big]\right\}
\frac{ \sin\big[F_*^{}(\lambda_{2}^{} -
\lambda_{1}^{})\big]}{\lambda_{2}^{} - \lambda_{1}^{}}
\nonumber \\
&~& + 4 {\rm Im}[\xi^{\alpha \beta }_2 \xi^{\alpha \beta*}_3]
\cos\big[F_*^{}( \lambda_{2}^{} - \lambda_{1}^{})\big] \frac{ \sin\big[F_*^{}( \lambda_{2}^{} - \lambda_{1}^{})\big]}{\lambda_{2}^{} - \lambda_{1}^{}}\; ,
\label{eq:Pab_full}
\end{eqnarray}
where the identity $\lambda^{}_1 + \lambda^{}_2 = -(b+\lambda^{}_3)$ has been implemented. Although we will not show the exact expressions of $\xi^{\alpha \beta}_i$'s, some useful properties of them can be implemented to further simplify the oscillation probabilities and the series expansions of $\xi^{\alpha \beta}_i$'s with respect to the small parameter $\alpha$ have been collected in Appendix A.

In the appearance channel $\nu^{}_\alpha \to \nu^{}_\beta$ with $\alpha \neq \beta$, the identity $\xi^{\alpha \beta}_1 = - \xi^{\alpha \beta}_2$ holds exactly. Therefore, it is easy to verify that ${\rm Im}[\xi^{\alpha \beta *}_1 \xi^{\alpha \beta}_2] = 0$, ${\rm Re}[\xi^{\alpha \beta *}_1 \xi^{\alpha \beta}_2] = -|\xi^{\alpha \beta}_1|^2 = - |\xi^{\alpha \beta}_2|^2$ and ${\rm Im}[\xi^{\alpha \beta *}_2 \xi^{\alpha \beta}_3] = - {\rm Im}[\xi^{\alpha \beta *}_1 \xi^{\alpha \beta}_3]$. Under these conditions, Eq.~(\ref{eq:Pab_full}) will be reduced to
\begin{eqnarray}
\widetilde{P}^{}_{\alpha \beta} &=&  \left\{4 |\xi_3^{\alpha \beta}|^2 - (\lambda_{2}^{} - \lambda_{1}^{})^{2} |\xi_1^{\alpha \beta}|^2 \right\}
\frac{\sin^2\big[F_*^{}(\lambda_{2}^{} - \lambda_{1}^{})\big]}{(\lambda_{2}^{} - \lambda_{1}^{})^2}
\nonumber \\
&~& +  2|\xi^{\alpha \beta}_1|^2 \left\{ 1 -
\cos\big[F_*^{} ( 3 \lambda_3^{} + b  ) \big] \cos \big[F_*^{}(\lambda_{2}^{} -
\lambda_{1}^{})\big]
\right\}\nonumber \\
&~& -4{\rm Re}[\xi^{\alpha \beta }_1 \xi^{\alpha \beta*}_3]
\sin\big[F_*^{} ( 3 \lambda_3^{} + b  ) \big]
\frac{ \sin\big[F_*^{}(\lambda_{2}^{} -
\lambda_{1}^{})\big]}{\lambda_{2}^{} - \lambda_{1}^{}}
\nonumber \\
&~& + 4 {\rm Im}[\xi^{\alpha \beta }_1 \xi^{\alpha \beta*}_3] \left\{
\cos\big[F_*^{}( 3 \lambda_3^{} + b  ) \big] -
\cos\big[F_*^{}( \lambda_{2}^{} - \lambda_{1}^{})\big] \right\} \frac{ \sin\big[F_*^{}( \lambda_{2}^{} - \lambda_{1}^{})\big]}{\lambda_{2}^{} - \lambda_{1}^{}}\; ,
\label{eq:Pab_app}
\end{eqnarray}
where one can observe four different types of oscillation terms. In the disappearance channel $\nu^{}_\alpha \to \nu^{}_\alpha$, we have $\xi^{\alpha \beta}_1 + \xi^{\alpha \beta}_2 = 1$ and $\xi^{\alpha \beta *}_i = \xi^{\alpha \beta}_i$, and thus arrive at
\begin{eqnarray}
\widetilde{P}^{}_{\alpha \beta} &=& 1 + \left\{4 |\xi_3^{\alpha \beta}|^2 - (\lambda_{2}^{} - \lambda_{1}^{})^{2} |\xi_2^{\alpha \beta}|^2 \right\}
\frac{\sin^2\big[F_*^{}(\lambda_{2}^{} - \lambda_{1}^{})\big]}{(\lambda_{2}^{} - \lambda_{1}^{})^2}
\nonumber \\
&~& -  2\xi^{\alpha \beta}_1 \xi^{\alpha \beta}_2 \left\{ 1 -
\cos\big[F_*^{} ( 3 \lambda_3^{} + b  ) \big] \cos \big[F_*^{}(\lambda_{2}^{} -
\lambda_{1}^{})\big]
\right\}\nonumber \\
&~& -4\xi^{\alpha \beta}_1 \xi^{\alpha \beta}_3
\sin\big[F_*^{} ( 3 \lambda_3^{} + b  ) \big]
\frac{ \sin\big[F_*^{}(\lambda_{2}^{} -
\lambda_{1}^{})\big]}{\lambda_{2}^{} - \lambda_{1}^{}}
\; ,
\label{eq:Pab_dis}
\end{eqnarray}
in which only three oscillation terms survive. In order for the oscillation probabilities to be valid for arbitrary neutrino energies and baseline lengths, as we shall see later, it is important to always keep those oscillation terms not expanded at all.

\section{Analytical and Numerical Results}

So far, all the analytical results in the previous section are exact. In this section, we will first expand the eigenvalues $\lambda^{}_i$'s in terms of $\alpha$ and derive the approximate formulas of neutrino oscillation probabilities in the general $\eta$-gauge. Simple and compact formulas in the special case of $\eta = \cos^2 \theta^{}_{12}$ then emerge in an obvious way. As a by-product, the mapping between effective and fundamental mixing parameters is also obtained. Finally, numerical verifications are carried out to show high precisions of our analytical formulas, in comparison with the exact ones.

\subsection{Approximate Formulas}
\begin{figure}[!t]
\centerline{
\includegraphics[scale=1.1]{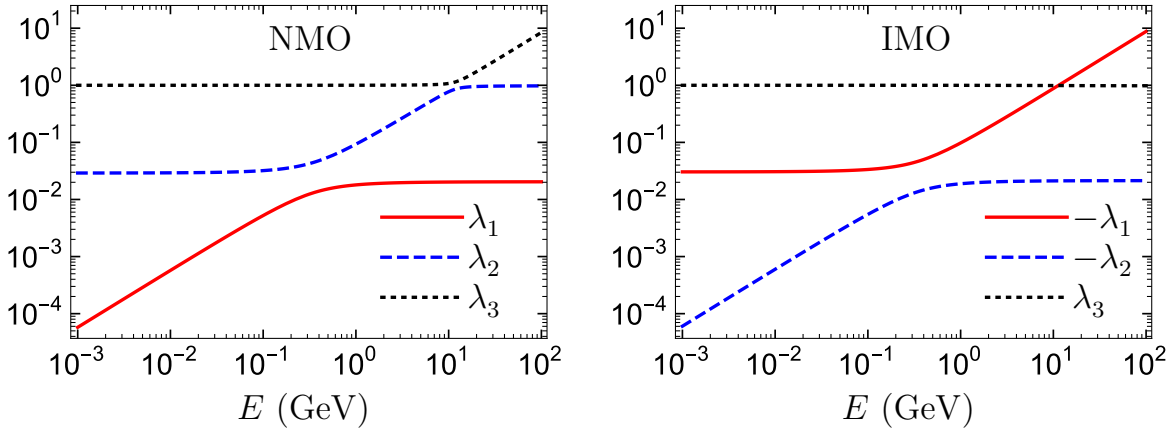}}
\caption{Three eigenvalues $\lambda_i^{}$ ($i=1,2,3$) of the matrix $M^{}_{\rm f}$ in Eq.~(\ref{eq:M_f}) shown as functions of the neutrino energy $E$, where the matter density $\rho \approx 2.8~{\rm g}~{\rm cm}^{-3}$, the electron fraction $Y^{}_e \approx 0.5$ and $\eta=1$ have been taken for illustration. The best-fit values of neutrino oscillation parameters from Table 1 have been adopted. The left panel is for the case of NMO while the right panel for IMO. Note that $\lambda^{}_1$ and $\lambda^{}_2$ are negative in the IMO case, so their absolute values have been plotted together with $\lambda^{}_3$ in the right panel.}
\label{fig:lambda}
\end{figure}
Let us begin with the series expansion of three eigenvalues. First, to clearly see the relative sizes of $\lambda^{}_i$'s, we have shown their exact values as functions of the neutrino energy $E$ in Fig.~\ref{fig:lambda}, where the matter density $\rho = 2.8~{\rm g}~{\rm cm}^{-3}$, the electron fraction $Y^{}_e \approx 0.5$ and $\eta=1$ have been taken for illustration. In addition, the best-fit values of neutrino oscillation parameters from Table 1 are adopted. In the left panel, the results for the NMO are given, where one can observe a potential level crossing at  $E \sim 0.3~{\rm GeV}$ for the solar resonance, and another one around $E \sim  10~{\rm GeV}$ for the atmospheric resonance, if a poor approximation to $\lambda_i^{}$ is adopted. In the right panel, since both $\lambda^{}_1$ and $\lambda^{}_2$ in the IMO case are actually negative, their absolute values are shown together with $\lambda^{}_3$. It is obvious that there is no level crossing in this case between $\lambda^{}_3$ and $\lambda^{}_{2}$, but  the level crossing at $E \sim 0.3~{\rm GeV}$ for the solar resonance still exists. For antineutrino oscillations in matter, as is well known, the atmospheric resonance will be present in the IMO case, while absent in the NMO case. A correct treatment of these eigenvalues in the regions of resonances is crucial to get well-behaved analytical results.

In this work we shall use $\alpha$ as the only expansion parameter and deal carefully with the would-be divergences in the neighborhood of resonances and in the limiting cases (e.g., the vacuum oscillations with $\widehat{A} \to 0$). Looking at the analytical results of $\lambda^{}_i$'s in Eqs.~(\ref{eq:lambda_NMO}) and (\ref{eq:lambda_IMO}), one should first expand $\sqrt{x^2 - 3 y} $ and $z$ in terms of $\alpha$, and then insert their approximate expressions back into Eqs.~(\ref{eq:lambda_NMO}) and (\ref{eq:lambda_IMO}). After a straightforward but tedious calculation, we finally get
\begin{eqnarray}
z & \approx &
\frac{1 + \widehat A + 3 \widehat C}{4 \widehat C'} + \frac{\alpha}{4 \widehat C
\widehat C'} \Big[2 \widehat C (1-2 \eta) - 3 (\eta - c^{2}_{\theta^{}_{12}}
) (1 - \widehat A c^{}_{2 \theta_{13}^{}} -\widehat C)\Big]
\nonumber\\ && +\frac{\alpha (1 + \widehat A + 3 \widehat C)}{8 (\widehat
 C')^3} \Big[ (2 \eta -1) (1+ \widehat A) -3 \widehat A  c^{2}_{\theta^{}
_{13}} (\eta - c^{2}_{\theta^{}_{12}}) \Big] \nonumber\\
&&- \frac{3 \alpha^2
\widehat A s^{2}_{2 \theta^{}_{12}} c^{2}_{\theta^{}_{13}} (1 - \widehat A
- \widehat C) }{4  \widehat C \widehat C' (1 + \widehat A + \widehat C)^2}
 + \frac{3 \alpha^2 \widehat A^2 s^{2}_{2\theta^{}_{13}}
(\eta - c^{2}_{\theta^{}_{12}} )^2 }{8 \widehat C^3 (\widehat C')}
\nonumber\\
&&- \frac{\alpha^2 }{8  \widehat C(\widehat C')^3} \Big[
2 \widehat C (1 - 2 \eta) - 3 (\eta -c^{2}_{\theta^{}_{12}} )
(1 - \widehat A c^{}_{2 \theta^{}_{13}} - \widehat C) \Big]
\nonumber\\ && \times
\Big[(1- 2 \eta) (1 + \widehat A) + 3 \widehat A c^{2}_{\theta^{}_{13}}
(\eta - c^{2}_{\theta^{}_{12}} )
\Big] -\frac{\alpha^2 (1 + \widehat A + 3 \widehat C)}{32 (\widehat C')^5}
\nonumber\\
&& \times \bigg\{ 4 \widehat (\widehat C')^2
 (1 - \eta + \eta^2) -3 \Big[
(1 - 2 \eta) (1 + \widehat A) + 3 \widehat A c^{2}_{\theta^{}_{13}}
(\eta - c^{2}_{\theta^{}_{12}} )\Big]^2
\bigg\} \;,
\label{eq:z}
\end{eqnarray}
and
\begin{eqnarray}
\sqrt{x^2 - 3 y} & \approx &
|\Delta_*^{}| \bigg\{ \widehat C' + \frac{\alpha}{2 \widehat C'}
\Big[ (1 - 2 \eta) (1 + \widehat A) +3 \widehat A c^{2}_{\theta^{}_{13}}
(\eta - c^{2}_{\theta^{}_{12}} ) \Big] \nonumber\\
&& + \frac{3 \alpha^2}{8 (\widehat C')^3} \Big[ \widehat C^2 + \widehat A
c^{2}_{\theta^{}_{13}} \big[s^{2}_{2 \theta^{}_{12}}
- 2 c^{}_{2 \theta_{12}^{} } (\eta - c^{2}_{\theta^{}_{12}}
) (1 - \widehat A) \nonumber\\
&&+ \widehat A (4 - 3 c^{2}_{\theta^{}_{13}})
(\eta - c^{2}_{\theta^{}_{12}})^2 \big]\Big]
\bigg\}     \;,
\label{eq:xy}
\end{eqnarray}
where  $c_{\phi}^{} \equiv \cos\phi$ and $s_{\phi}^{} \equiv \sin \phi$ have been introduced also for $\phi = 2\theta^{}_{ij}$. In addition, we have defined a regulator for the atmospheric resonance~\cite{Freund:2001pn}
\begin{eqnarray}
\widehat C & = &
\sqrt{(1 - \widehat A)^2 + 4 \widehat A s^{2}_{\theta^{}_{13}}} \; ,
\label{eq:C_hat}
\end{eqnarray}
and $\widehat C' \equiv (\widehat C^2 + \widehat A c^2_{\theta^{}_{13}})^{1/2}$. Note that $\widehat C$ appears in the denominators and will cause divergences in the further expansions in terms of $\sin^2\theta_{13}$ when $\widehat A=1$. Therefore, we shall keep the exact form of $\widehat C$ in Eq.~(\ref{eq:C_hat}) in our calculations of the oscillation probabilities.

On the other hand, in the low-energy or vacuum limit with $\widehat{A} \to 0$, we have learned from Refs.~\cite{Xu:2015kma} and \cite{Xing:2016ymg} that one cannot expand the function $\widehat{\epsilon}$ mentioned in Section 1 in terms of $\alpha$. A further study shows that this function arises from the difference between two eigenvalues $\lambda^{}_2$ and $\lambda^{}_1$, namely, the terms proportional to $\sqrt{1 - z^2}$ in Eqs.~(\ref{eq:lambda_NMO}) and (\ref{eq:lambda_IMO}). Therefore, we define $\epsilon \equiv \lambda^{}_2 - \lambda^{}_1$ and expand $\lambda^{}_3$ up to the second order of $\alpha$. Then, $\lambda^{}_1$ and $\lambda^{}_2$ can be obtained from the identity $\lambda^{}_1 + \lambda^{}_2 = -(b+\lambda^{}_3)$ and the definition of $\epsilon$, namely,
\begin{eqnarray}
\lambda_1^{} & \approx &
-\frac{1}{2}(b + \rho_1^{}+ \rho_2^{} \alpha + \rho_3^{} \alpha^2
+\epsilon)
\;,
\nonumber\\
\lambda_2^{} & \approx &
-\frac{1}{2}(b + \rho_1^{}+ \rho_2^{} \alpha + \rho_3^{} \alpha^2
-\epsilon) \;,
\nonumber\\
\lambda_3^{} & \approx &
\rho_1^{} + \rho_2^{} \alpha +  \rho_3^{} \alpha^2 \; ,
\label{eq:lambda_app}
\end{eqnarray}
where the higher-order terms of ${\cal O}(\alpha^3)$ have been omitted. Note that Eq.~(\ref{eq:lambda_app}) is valid for both NMO and IMO. The corresponding coefficients $\rho^{}_i$ (for $i = 1, 2, 3$) in Eq.~(\ref{eq:lambda_app}) can be directly computed by making use of Eqs.~(\ref{eq:bcd}), (\ref{eq:z}) and (\ref{eq:xy}). More explicitly, we have
\begin{eqnarray}
\rho^{}_1  & = &
\frac{1 + \widehat A + \widehat C}{2}  \; ,\nonumber\\
\rho^{}_2  & = & \displaystyle
\frac{( \eta- c^{2}_{\theta^{}_{12}}
)  \left[ -1 + \widehat C + \widehat A c^{}_{2 \theta^{}_{13}}
) \right]}{2 \widehat C} \; ,\nonumber\\
\rho^{}_3  & = & \displaystyle
( \eta - c^{2}_{\theta^{}_{12}} )^2
s^{2}_{2 \theta^{}_{13}} \frac{\widehat A^2}{4 \widehat C^3}
-\frac{s^{2}_{2\theta^{}_{12}} ( 1 - \widehat A - \widehat C)
( 1 + \widehat A - \widehat C)}
{8 \widehat C ( 1 + \widehat A + \widehat C)}
\; ,
\label{eq:rhoi}
\end{eqnarray}
where one can clearly observe that the above coefficients will be greatly simplified for $\eta = \cos^2 \theta^{}_{12}$. In particular, $\rho^{}_2 = 0$ implies that the first order correction to $\lambda^{}_3$ is vanishing, so the leading-order results are already very precise. Additionally, at the second order, only one term is left in $\lambda^{}_3$. However, this is not the case for $\lambda^{}_1$ and $\lambda^{}_2$, as extra contributions come from $\epsilon$, which can be determined from
\begin{eqnarray}
\epsilon^2  & \approx &
\frac14 \biggm\{ 1 + \widehat A - \widehat C + 2 \alpha \Bigr[
2 \eta -1 + \frac{ (\eta - c^{2}_{\theta^{}_{12}})
( 1 - \widehat A c^{}_{2 \theta^{}_{13}} - \widehat C)}{2 \widehat C} \Bigr]
\biggm\} ^2 \nonumber\\ && -2 \alpha (1 + \widehat A -\widehat C) (\eta +
c^{2}_{\theta^{}_{12}} -1) + \frac{2 \alpha^2 \widehat A^2
(1 - \widehat A -\widehat C)}{ \widehat C (1 + \widehat A + \widehat C)^3}
c^{4}_{\theta^{}_{13}} s^{2}_{2 \theta^{}_{12}}
\nonumber\\ &&
- \frac{8 \alpha^2(1 + \widehat A s^{2}_{\theta^{}_{13}}) (\eta -1) \eta}{
1 + \widehat A + \widehat C}- \frac{4 \alpha^2 \widehat A^3
(\eta -c^{2}_{\theta^{}_{12}})^2}{ \widehat C^3 (1 + \widehat A +\widehat C) }
c^{4}_{ \theta^{}_{13}} s^{2}_{\theta^{}_{13}}
\nonumber\\ &&
- \frac{8 \alpha^2
\widehat A (\eta - c^{2}_{\theta^{}_{12}})}{\widehat C (1 + \widehat A
+ \widehat C)^2} (\eta + c^{2}_{\theta^{}_{12}} -1) (1 - \widehat A c^{}_{2
\theta^{}_{13}} - \widehat C)c^{2}_{\theta^{}_{13}} \; .
\label{eq:epsilon}
\end{eqnarray}
In the derivation of Eq.~(\ref{eq:epsilon}), the identities $\lambda^{}_1 + \lambda^{}_2 = -(b+\lambda^{}_3)$ and $\lambda^{}_1 \lambda^{}_2 = -d\lambda^{-1}_3$ have been used, where both $b$ and $d$ have been given in Eq.~(\ref{eq:bcd}). Note that, instead of $\epsilon$ itself, $\epsilon^2$ has been expanded in $\alpha$ in Eq.~(\ref{eq:epsilon}) where high-order terms of ${\cal O}(\alpha^3)$ have been neglected. As we will show in the next subsection, $\epsilon$ reduces to $\widehat{\epsilon}$ in the case of $\eta = \cos^2 \theta^{}_{12}$ and in the limit of $\widehat{A} \to 0$. Therefore, $\epsilon$ is the parameter that we should retain in the series expansion.

Having obtained $\lambda^{}_i$'s, we can calculate $\xi_{i}^{ \alpha\beta}$'s according to Eqs.~(\ref{eq:CH}) and (\ref{eq:si}). Their analytical expressions have been collected in Appendix A.
After inserting $\xi_{i}^{ \alpha\beta}$'s and $\lambda^{}_i$'s into Eqs.~(\ref{eq:Pab_app}) and (\ref{eq:Pab_dis}), we finally obtain the approximate formulas of oscillation probabilities
\begin{eqnarray}
\widetilde P_{ee}^{} & \approx &
1 - 2 \Bigr[\frac{s^{2}_{2 \theta_{13}^{}} }{4 \widehat C^2} - \frac{\alpha\widehat A ( \eta -c^{2}_{\theta_{12}^{}} ) (
\widehat A - c^{}_{2\theta_{13}^{}})}
{2 \widehat C^4} s^{2}_{2 \theta_{13}^{}} \Bigr] (1 - \cos
\widetilde F_+^{} \cos\widetilde F_-^{})
\nonumber\\ &&
+\Bigr[ \frac{1 + \widehat A - \widehat C - 2 \alpha c^{}_{2\theta^{}_{12}}
}{4 \epsilon\widehat C^2 } s^{2}_{2\theta^{}_{13}}
- \frac{2 \alpha \widehat A (\eta - c^{2}_{\theta_{12}^{}} )
}{\epsilon \widehat C^4 ( 1 + \widehat A + \widehat C )}
\nonumber\\
&&\times(1 - 6 \widehat A c^{}_{2\theta_{13}^{}} - \widehat C + \widehat A \widehat C + 5
\widehat A^2)c^{4}_{\theta_{13}^{}} s^{2}_{\theta_{13}^{}}\Bigr]
\sin\widetilde F_+^{} \sin\widetilde F_-^{} \nonumber\\
&& -\frac{4 \alpha^2 ( 1 - \widehat A + \widehat C )}
{\epsilon^2\widehat C ( 1 + \widehat A + \widehat C )^3} s^{2}_{2\theta_{12}
^{}} c^{4}_{\theta_{13}^{}} \sin^2 \widetilde F_-^{} \; ,
\label{eq:Pee}
\end{eqnarray}
\begin{eqnarray}
\widetilde P_{\mu e}^{} & \approx& \Bigr[\frac{ s^2_{2\theta_{13}^{}} s^2_{\theta_{23}
^{}} }{2 \widehat C^2} - \frac{4 \alpha ( 1 - \widehat A - \widehat C
) }{\widehat C^2 ( 1 + \widehat A + \widehat C ) } {\cal J} \cot \delta -
\frac{\alpha\widehat A ( \eta -c^2_{\theta_{12}^{}}) ( \widehat A -
c^{}_{2 \theta_{13}^{}} )}{\widehat C^4} s^2_{2\theta_{13}^{}}
s^2_{\theta_{23}^{}} \Bigr]
\nonumber\\  &&
\times (1 - \cos \widetilde F_+^{} \cos
\widetilde F_-^{}) -\frac{2}{\epsilon}
\Bigr[\frac{( 1 + \widehat A - \widehat C - 2 \alpha c^{}_{2\theta^{}
_{12}})} {8 \widehat C^2 } s^2_{2
\theta_{13}^{}} s^2_{\theta_{23}^{}}
- \frac{\alpha ( 1 - \widehat A +
\widehat C)}{\widehat C^2} {\cal J} \cot \delta
\nonumber \\ &&
-\frac{\alpha \widehat A ( \eta -
c^2_{\theta_{12}^{}})}{\widehat C^4 ( 1 + \widehat A + \widehat C) }  (
1 - 6 \widehat A c^{}_{2\theta_{13}^{}} - \widehat C + \widehat A \widehat C + 5
\widehat A^2) c^4_{\theta_{13}^{}} s^2_{\theta_{13}^{}}
s^2_{\theta_{23}^{}} \Bigr]
 \nonumber\\ &&
\times\sin\widetilde F_+^{} \sin\widetilde F_-^{}
+\biggm\{\frac{\alpha^2 (1 - \widehat A + \widehat C)}{
\widehat C (1 + \widehat A + \widehat C )}
s^{2}_{2\theta_{12}^{}} c^{2}_{\theta_{13}^{}}
c^{2}_{\theta_{23}^{}}
+ \frac{16 \alpha {\cal J \cot \delta}}{\widehat C
(1 + \widehat A + \widehat C )^2}\Bigr[ \alpha c^{}_{ 2\theta_{12}^{}}
\nonumber \\ && \times
(\widehat C + \widehat A c^2 _{\theta_{13}^{}}) - \widehat A
c^2 _{\theta_{13}^{}}\Bigr]
- \frac{\alpha^2
(1 + \widehat A)}{\widehat C(1 + \widehat A + \widehat C)^2}
s^{2}_{2\theta_{12}^{}} s^{2}_{2 \theta_{13}^{}}
s^{2}_{\theta_{23}^{}}
\nonumber\\ && +
\frac{16 \alpha^2 \widehat A ( \eta -c^2_{\theta_{12}^{}})}
{\widehat C^3 ( 1 + \widehat A + \widehat C)^2} ( 1 - 3 \widehat A c^{}
_{2\theta_{13}^{}} - \widehat C + \widehat A \widehat C + 2 \widehat A^2 )
c^2_{\theta_{13}^{}} {\cal J } \cot\delta\biggm\}\frac{\sin^2
\widetilde F_-^{}}{\epsilon^2}\nonumber\\ &&
+ \frac{8 \alpha {\cal J}}{\epsilon \widehat C (1 + \widehat A +\widehat C)}
(\cos \widetilde F^{}_{+} - \cos \widetilde F^{}_{-}
) \sin \widetilde F^{}_{-} \;,
\label{eq:Pme}
\end{eqnarray}
where
\begin{eqnarray}
\widetilde F_{-}^{} & = &  \epsilon F_{*}^{}\;,
\nonumber \\
\widetilde F_{+}^{}  & = & \displaystyle \frac{F_{*}^{}
}{2}\Bigr[1 + \widehat A + 3 \widehat C - 2\alpha (2 \eta -1)
-\frac{3 \alpha (\eta - c^2_{\theta^{}_{12}})
(1 - \widehat A  c^{}_{2 \theta^{}_{13}} - \widehat C)}{\widehat C}\Bigr]
\;.
\label{eq:Fpm}
\end{eqnarray}
The expansion of $\widetilde{F}^{}_+$ is given to ${\cal O}(\alpha)$, but a few terms proportional to $\alpha^2$ are kept in the coefficients in front of oscillation terms, as they may become important in some cases. For completeness, we also present the complete expression for $\widetilde{P}_{\tau\mu}^{}$ in Eq.~(56) in Appendix B. As is proved in Ref.~\cite{Akhmedov:2004ny}, only two oscillation probabilities are independent, say, $\widetilde{P}^{}_{\mu e}$ and $\widetilde{P}^{}_{\tau \mu}$. The other probabilities can be constructed by making use of unitarity condition $\sum_\alpha \widetilde{P}^{}_{\alpha \beta} = \sum_\beta \widetilde{P}^{}_{\alpha \beta} = 1$ and the time-reversal transformation $\widetilde{P}^{}_{\alpha \beta} = \widetilde{P}^{}_{\beta\alpha}(\delta \to -\delta)$ for a constant matter density. Furthermore, considering that the rotation matrix in the $2$-$3$ sector commutes with the matter potential term in the effective Hamiltonian, we can establish the relations $\widetilde{P}^{}_{e\tau} = \widetilde{P}^{}_{e\mu}(\theta^{}_{23} \to \theta^{}_{23} + \pi/2)$ and $\widetilde{P}^{}_{\mu\mu} = \widetilde{P}^{}_{\tau\tau}(\theta^{}_{23} \to \theta^{}_{23} + \pi/2)$. For this reason, only two independent appearance probabilities $\widetilde{P}^{}_{\mu e}$ and $\widetilde{P}^{}_{\tau \mu}$ are shown in this work, while $\widetilde{P}^{}_{ee}$ is given as an example in the disappearance channel.  Note that the oscillation probabilities are for NMO, and we can get the corresponding results by replacing $\epsilon$ with $- \epsilon$ for IMO.

Given the above approximate formulas for oscillation probabilities, we also verify that these expansions indeed reduce to those that already exist in the literature. For example, in the low energy range, $\alpha$, $\widehat{A}$ and $\epsilon$ are of the same order and can be expanded simultaneously, from which one can arrive at Eq.~(4.6) in Ref.~\cite{Xing:2016ymg}. On the other hand, for the high energy region with $\widehat{A}\sim\epsilon\gg\alpha$, one can safely
expand $\epsilon$ in terms of $\alpha$ and restore the familiar results of Freund~\cite{Freund:2001pn} and Akhmedov {\it et al.}~\cite{Akhmedov:2004ny}.

As we have mentioned, the analytical expressions can be substantially simplified when $\eta=\cos^2 \theta^{}_{12}$ is adopted, where all the terms proportional to $(\eta-\cos^2 \theta^{}_{12})$ automatically disappear. The resultant simplified formulas will be presented and discussed in the next subsection. Here we show that the choice of
$\eta=\cos^2 \theta^{}_{12}$ is not only advantageous for the analytical simplicity, but also for numerical accuracy.
\begin{figure}
\centerline{
\hspace{-0.3cm}
\includegraphics[scale=0.88]{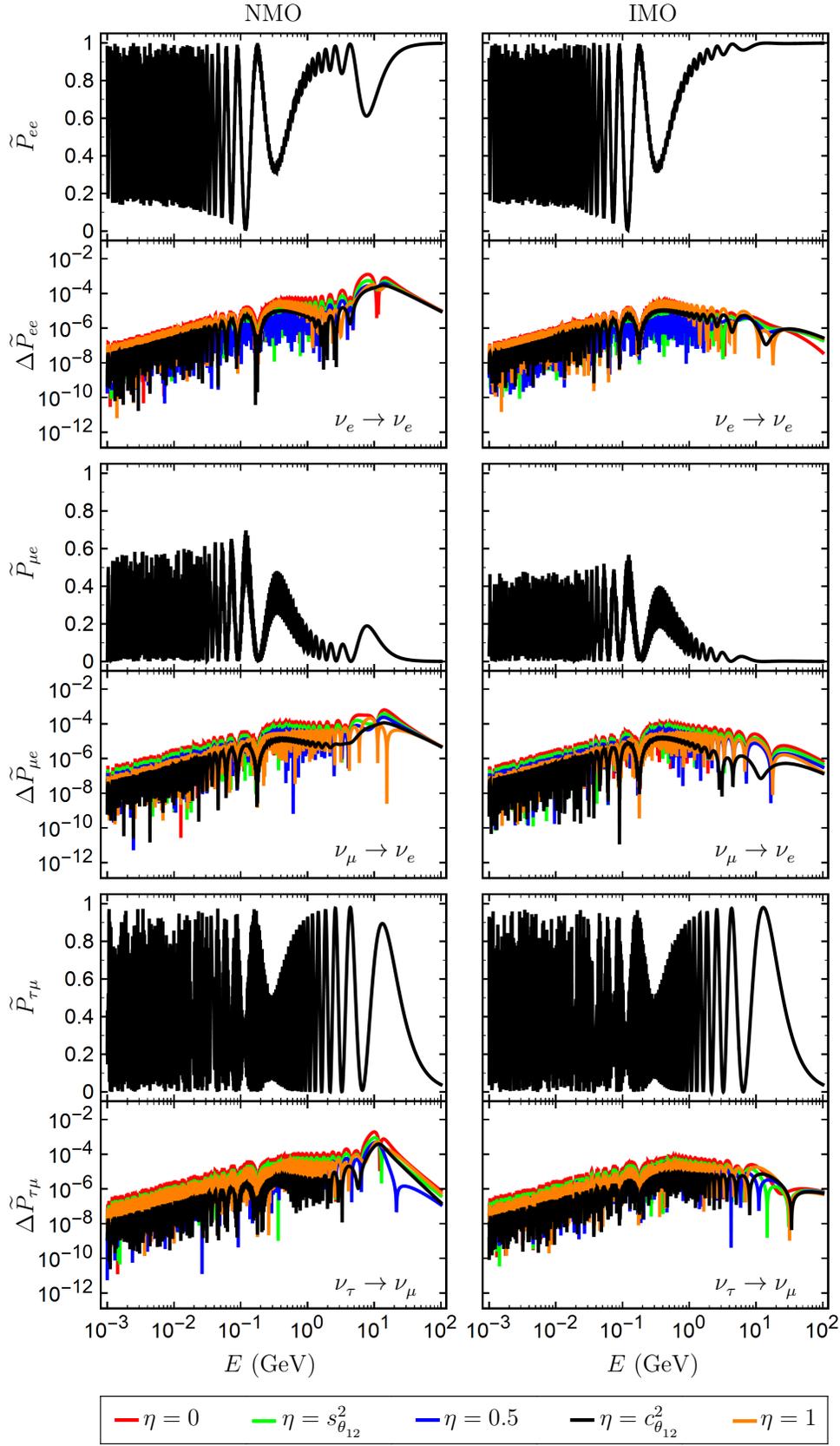}}
\caption{Accuracy tests of the analytical approximations of neutrino oscillation probabilities $\widetilde{P}^{}_{\alpha \beta}$ for different choices of $\eta$. The best-fit values of neutrino oscillation parameters in Table~1 have been adopted and a baseline of $L =6500~ {\rm km}$ is employed. The left panel is for NMO and the right panel is for IMO.}
\label{fig:error}
\end{figure}
To examine the influence of $\eta$ on the accuracy of analytical approximations in the oscillation probabilities, we have adopted the best-fit values of neutrino oscillation parameters in Table 1. In addition, the matter density $\rho \approx 2.8 ~{\rm g}~{\rm cm}^{-3}$ for the
Earth's crust and $Y^{}_e \approx 0.5$ are taken for illustration. In order to test the numerical accuracy, we define the absolute error of the analytical approximations of $\widetilde
P (\nu_{\alpha} \to \nu_{\beta})$ as $\Delta \widetilde
P _{\alpha \beta}$ for $\alpha,~\beta =e ,~\mu,~\tau$, i.e.,
\begin{eqnarray}
\Delta \widetilde P _{\alpha \beta} =  |(\widetilde P _{\alpha
\beta})^{}_{\rm Exact} - (\widetilde P _{\alpha
\beta})^{}_{\rm Approximate}| \; ,
\label{eq:DeltaP}
\end{eqnarray}
where $(\widetilde P _{\alpha \beta})^{}_{\rm Exact}$ is calculated by a fully numerical evolution of the neutrino flavor states.
Note that an unusual baseline of $L=6500$ km is employed in order to make the fine structure of oscillations more prominent.
The oscillation probabilities and their absolute errors are given in Fig.~\ref{fig:error}, where we can observe that the case of $\eta=\cos^2 \theta^{}_{12}$ is the most accurate one for almost the entire range of neutrino energies.\footnote{Note that the spikes along the curves for $\Delta \widetilde{P}^{}_{\alpha \beta}$ in Fig.~\ref{fig:error} do not mean the best precision but the intersection points of exact and approximate oscillation probabilities, which are caused by the modifications of oscillation frequency and amplitude in the approximate formulas.}

Comparing with previous analytical approximations of the oscillation probabilities, our results are advantageous in several aspects.
First, we have included all the possible leading terms of the whole energy region. Taking the expansion terms $\alpha^2/\epsilon^2$ and $\alpha$ for instance, although $\alpha^2/\epsilon^2$ is a higher-order term than $\alpha$ near the atmospheric resonance, it is significantly enhanced in the low energy range where $\epsilon$ is small. Thus both are maintained in the expansion. Second, our analytical results keep $\epsilon$ and $\widehat C$ as independent parameters in order to avoid any divergence in the low-energy limit and near the atmospheric resonance, respectively. Third, for the first time, we have presented the analytical results with a generic $\eta$ value, which is convenient to make a comparison with previous results. We further show that $\eta=\cos^2 \theta^{}_{12}$ is the best choice in terms of both simplicity and numerical accuracy.\footnote{Although we demonstrate that $\eta = \cos^2 \theta^{}_{12}$ leads to simpler and more accurate oscillation probabilities, the underlying physical reason is not clear and deserves further studies~\cite{Zhou:2016luk}. We notice that the same mass-squared difference $\Delta m^2_{ee} \equiv \cos^2\theta^{}_{12} \Delta^{}_{31} + \sin^2\theta^{}_{12} \Delta^{}_{32}$ has been shown in Ref.~\cite{Parke:2016joa} to be advantageous for $\overline{\nu}^{}_e$ disappearance experiments without matter effects. This observation may provide a clue to better understand the choice of $\eta = \cos^2 \theta^{}_{12}$.}

\subsection{Special Case of $\eta = \cos^2\theta^{}_{12}$}
If $\eta = \cos^2 \theta^{}_{12}$ is fixed, we can obtain much simpler formulas for relevant oscillation parameters in matter and those for the oscillation probabilities as well. First, let us focus on the two regulators for eliminating possible divergences. As indicated in Eq.~(\ref{eq:C_hat}), $\widehat{C}$ depends on $\eta$ implicitly through $\widehat{A}$ and $\Delta^{}_*$, so its expression is not modified. The other one is
\begin{eqnarray}
\epsilon & \approx & \displaystyle \sqrt{\frac{  (1 + \widehat A - \widehat C -2 \alpha c_{ 2 \theta_{12}^{}}^{}  )^2}{4}  + \frac{2   \alpha^2 (1+\widehat A s_{\theta_{13}}^2)s_{2 \theta_{12}^{}}^{2}}{1 + \widehat A + \widehat C}^{} + \frac{2 \alpha^2 \widehat A^2 (1 - \widehat A - \widehat C) c_{\theta_{13}^{}}^4 s^{2}_{2 \theta_{12}^{}}}{\widehat C (1 + \widehat A + \widehat C)^3}} \; , \quad
\label{eq:epsilon_new}
\end{eqnarray}
where Eq.~(\ref{eq:epsilon}) with $\eta = \cos^2 \theta^{}_{12}$ has been used.

In the low-energy limit, $\widehat{A}$ will also be a small parameter, just like $\alpha$. In this case, it is easy to verify
\begin{eqnarray}
\epsilon & \approx & \widehat{\epsilon} \equiv \sqrt{\alpha^2 +  \widehat A^2 c^{4}_{\theta^{}_{13}} -2 \alpha \widehat A c^{}_{2 \theta^{}_{12}} c^{2}_{\theta^{}_{13}} } \; ,
\label{eq:epsilon_low}
\end{eqnarray}
where higher-order terms of ${\cal O}(\alpha^2\widehat{A})$ are omitted. It has been found in Ref.~\cite{Xing:2016ymg} that one can keep $\epsilon$ in the oscillation probabilities, whose low-energy behaviors will then be remarkably improved. In the high-energy limit, it is safe to expand $\epsilon$ in terms of $\alpha$, and thus we get
\begin{eqnarray}
\epsilon \approx \frac{1+\widehat{A} - \widehat{C} - 2 \alpha c_{2\theta_{12}^{}}^{}}{2} + \frac{\alpha^2 (1 - \widehat A + \widehat C) (1 + \widehat A + \widehat C)
}{8 \widehat C (1 + \widehat A - \widehat C)} s^2_{2 \theta^{}_{12}}
+\frac{\alpha^2 (1 + \widehat A s^2_{\theta^{}_{13}})}{4 c^2_{\theta^{}
_{13}} \widehat A } s^2_{2 \theta^{}_{12}} \; .
\label{eq:epsilon_hi}
\end{eqnarray}
Note that $\epsilon$ in this case is not a small parameter, as the matter effects become important or even dominant, e.g., $\widehat{A} \gtrsim 1$. For an arbitrary neutrino energy, it is necessary to make use of the full result of $\epsilon$ in Eq.~(\ref{eq:epsilon_new}).

Second, as it is useful to define the oscillation phases $F^{}_- \equiv \Delta^{}_{21}L/(4E)$ and $F^{}_+ \equiv (\Delta^{}_{31} + \Delta^{}_{32})L/(4E)$ in vacuum, we obtain the following analytical approximations of their counterparts in matter with the help of Eq.~(\ref{eq:Fpm}), namely,
\begin{eqnarray}
\widetilde F_{-}^{} = \epsilon F_{*}^{}
 \;, \quad
\widetilde F_{+}^{}  =  \frac{F_{*}^{}
}{2} (1 + \widehat A - 2\alpha c^{}_{ 2\theta_{12}^{}} + 3 \widehat C) \; ,
\label{eq:Fpm_mat}
\end{eqnarray}
reflecting the corrections induced by the Earth matter to neutrino mass-squared differences.

Given the above parameters, the oscillation probabilities for the special case of $\eta = c^2_{\theta^{}_{12}}$ turn out to be
\begin{eqnarray}
\widetilde P^{}_{ee}  & \simeq & 1 - \frac{s^2_{2 \theta_{13}^{}}
}{2 \widehat C^2}  (1 - \cos\widetilde F_{+}^{} \cos\widetilde F_{-}^{}
 ) + \frac{s^2_{2 \theta_{13}^{}}}{4
\epsilon \widehat C^2}  (1 + \widehat A - \widehat C - 2\alpha c^{}_{2 \theta_{12}^{}}
 )\sin\widetilde F_{+}^{} \sin\widetilde F_{-}^{}
\nonumber \\
&& -\frac{ 4 \alpha^2  (1 - \widehat A + \widehat C) }{\epsilon^2 \widehat C
 (1 + \widehat A + \widehat C)^3 } s^{2}_{2\theta_{12}^{}}c^{4}_{\theta_{13}^{}}
\sin^2 \widetilde F_{-}^{} \; ,
\label{eq:Pee_new}
\end{eqnarray}
and
\begin{eqnarray}
\widetilde P^{}_{\mu e} \hspace{-0.15cm} & \simeq &
\hspace{-0.15cm}
\Bigr [ \frac{s^{2}_{2 \theta_{13}^{}} s^{2}_{\theta_{23}^{}}
}{2 \widehat C^2} - \frac{4 \alpha  (1 - \widehat A - \widehat C)}
{\widehat C^2  (1 + \widehat A + \widehat C)}  {\cal J} \cot\delta
\Bigr ]   (1 - \cos\widetilde F_{+}^{} \cos\widetilde F_{-}^{}
 ) \nonumber \\
&& -  \Bigr[\frac{ s^{2}_{2 \theta_{13}^{}}
s^{2}_{\theta_{23}^{}} } {4 \epsilon \widehat C^2}
 (1 + \widehat A - \widehat C -2 \alpha c^{}_{2\theta_{12}^{}} ) - \frac{2 \alpha  (1 - \widehat A + \widehat C)}
{\epsilon \widehat C^2} {\cal J} \cot\delta \Bigr ]
\sin\widetilde F_{+}^{} \sin\widetilde F_{-}^{}
\nonumber \\
&& +   \biggm\{ \frac{\alpha^2  (1 - \widehat A + \widehat C)}{
\widehat C  (1 + \widehat A + \widehat C  )}
s^{2}_{2\theta_{12}^{}} c^{2}_{\theta_{13}^{}}
c^{2}_{\theta_{23}^{}}
- \frac{\alpha^2
(1 + \widehat A)}{\widehat C(1 + \widehat A + \widehat C)^2}
s^{2}_{2\theta_{12}^{}} s^{2}_{2 \theta_{13}^{}}
s^{2}_{\theta_{23}^{}}
 \nonumber \\ \hspace{-0.15cm} &  &
\hspace{-0.15cm} \displaystyle
+ \frac{16 \alpha {\cal J \cot \delta}}{\widehat C
 (1 + \widehat A + \widehat C  )^2}  \Bigr[ \alpha c^{}_{ 2\theta_{12}
^{}}  (\widehat C + \widehat A c^2 _{\theta_{13}^{}} ) - \widehat A
c^2 _{\theta_{13}^{}} \Bigr]   \biggm\}
\frac{\sin^2 \widetilde F_{-}^{}} {\epsilon^2}
\nonumber \\
&& - \frac{8 \alpha {\cal J} }{\epsilon \widehat C  (1 + \widehat A + \widehat C
 )}  (\cos \widetilde F^{}_{-} - \cos \widetilde F^{}_{+}
 ) \sin \widetilde F^{}_{-} \; ,
 \label{eq:Pme_new}
\end{eqnarray}
which are much simpler than the general formulas in Eqs.~(\ref{eq:Pee}) and (\ref{eq:Pme}). See also the results of $\widetilde{P}^{}_{\tau \mu}$ in Eq.~(57) in Appendix B.

\begin{figure}[H]
\centerline{
\includegraphics[scale=0.9]{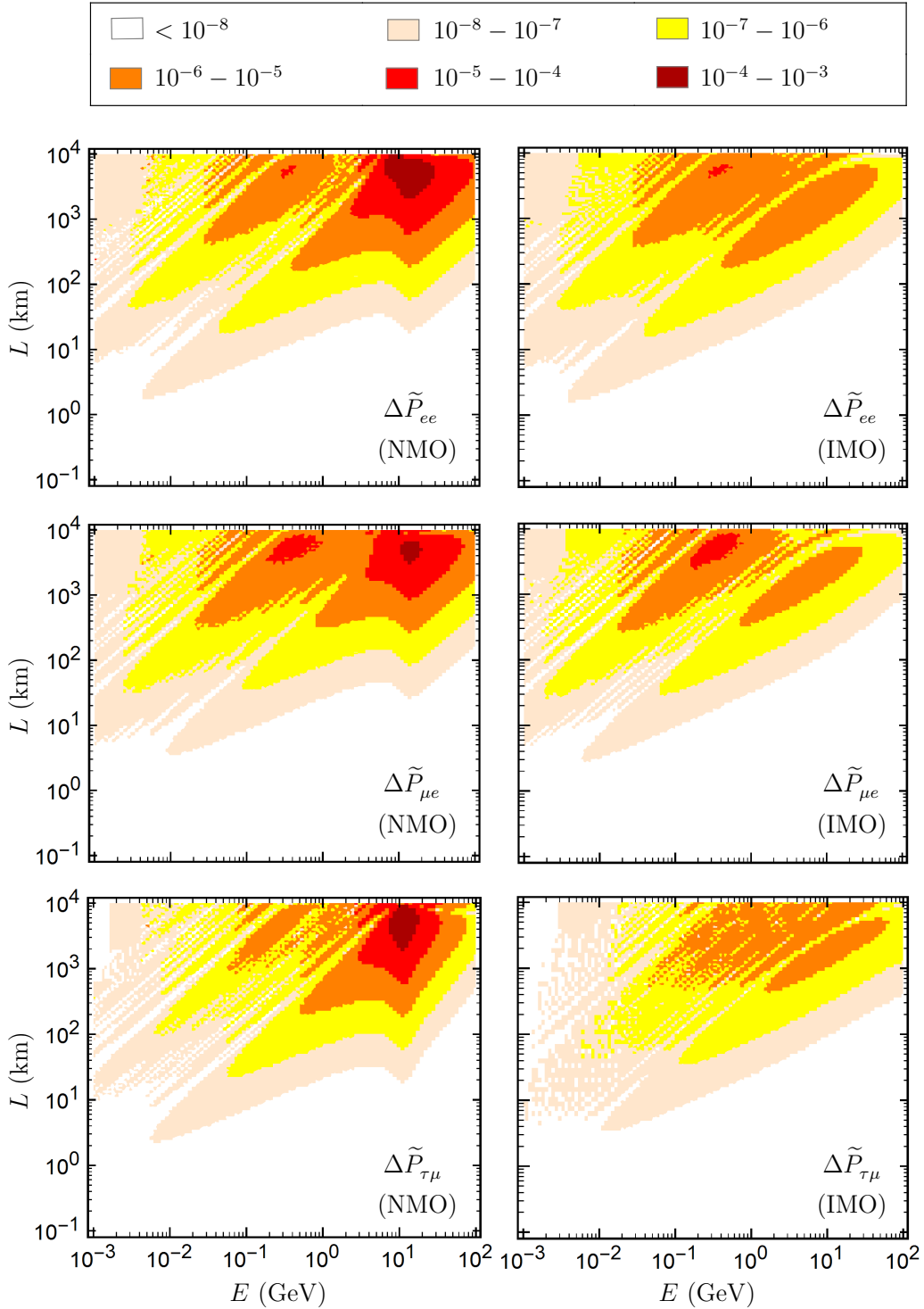}}
\caption{ The accuracy tests of analytical approximations
to neutrino oscillation probabilities for $\eta = c^2_{\theta^{}_{12}}$, where the matter density of $\rho \approx 2.8~{\rm g}~{\rm cm}^{-3}$ with the electron fraction $Y^{}_e \approx 0.5$ and the best-fit values of neutrino oscillation parameters in Table~1 have been used. The absolute errors $\Delta \widetilde{P}^{}_{\alpha \beta}$ (for $\alpha \beta = ee, \mu e, \tau \mu$) have been defined in Eq.~(\ref{eq:DeltaP}), and the probabilities are averaged over a Gaussian energy resolution of $1\%$. }
\label{fig:2D}
\end{figure}

To carry out a systematic test of numerical accuracy of analytical approximations, we consider the absolute errors $\Delta \widetilde{P}^{}_{\alpha \beta}$ defined in Eq.~(\ref{eq:DeltaP}) and the approximate results are now obtained by using the simplified formulas in the case of $\eta = \cos^2 \theta^{}_{12}$. The numerical results of $\Delta \widetilde{P}^{}_{\alpha \beta}$ for a wide range of neutrino energies and baseline lengths have been shown in Fig.~\ref{fig:2D}, where the sizes of absolute errors are denoted by different colors. Some comments on the numerical calculations are in order:
\begin{itemize}
\item In Fig.~\ref{fig:2D}, the matter density of $\rho \approx 2.8~{\rm g}~{\rm cm}^{-3}$ with $Y^{}_e \approx 0.5$ and the best-fit values of neutrino oscillation parameters from Table~1 have been used in numerical calculations. In addition, to avoid fast oscillations at low energies, we have averaged the oscillation probabilities over a Gaussian energy resolution of $1\%$. The baseline lengths and neutrino energies have been set to be $0.1~{\rm km} \leq L \leq 10^4~{\rm km}$ and $1~{\rm MeV} \leq E \leq 100~{\rm GeV}$, respectively. Hence, both current and future oscillation experiments as mentioned in the introduction are essentially covered. As for the atmospheric neutrinos, our assumption of a constant matter density renders it impossible to reveal the structure of parametric resonances~\cite{Liu:1997yb,Liu:1998nb,Petcov:1998su,Chizhov:1999az,Chizhov:1999he}. However, it suffices to illustrate the numerical difference between our analytical formulas and the exact oscillation probabilities.

\item In the lower part of each plot in Fig.~\ref{fig:2D}, i.e., for $L \leq 1~{\rm km}$, one can observe that the errors are always far below the level of $10^{-8}$. This can be understood by noticing that the oscillations driven by $\Delta^{}_{21}$ have not yet developed for a short baseline. The $\Delta^{}_{31}$-driven oscillations indeed take place for short baseline lengths and low neutrino energies, however, the amplitudes will be suppressed by the smallest mixing angle $\theta^{}_{13}$. For higher neutrino energies, we need longer baseline lengths for the $\Delta^{}_{31}$-driven oscillations to develop. The errors in the entire range of baseline lengths and energies are below $10^{-3}$, demonstrating an excellent agreement between our approximate formulas and the exact ones.

\item For IMO on the right column of Fig.~\ref{fig:2D}, one can observe that the discrepancy is at most $10^{-5} \sim 10^{-4}$, as a consequence of the absence of resonances in this case. For NMO, the region of largest errors always appears around $E \approx 10~{\rm GeV}$ and $L \approx 5000~{\rm km}$, where the atmospheric resonance is encountered, while around the region of the solar resonance relatively smaller errors are observed. Such a difference on the size of error at two different resonances may be attributed to the fact that $\lambda_2^{}$ and $\lambda_3^{}$ are more close to each other at the atmospheric resonance than $\lambda_1^{}$ and $\lambda_2^{}$ at the solar resonance.
\end{itemize}
Notice that the approximate formulas of $\widetilde{P}^{}_{ee}$ and $\widetilde{P}^{}_{\mu e}$ in Eqs.~(\ref{eq:Pee_new}) and (\ref{eq:Pme_new}), together with that of $\widetilde{P}^{}_{\tau \mu}$ in Eq.~(57) in Appendix B, are the main results of this work. Given their simplicity and high level of numerical accuracy, one may directly employ them to perform both analytical and numerical studies on neutrino oscillation phenomena in current and upcoming oscillation experiments. We leave such applications for a future work.

\subsection{Parameter Mapping}

With those newly obtained approximate formulas for oscillation probabilities, a more accurate mapping of the intrinsic mixing parameters to the effective mixing parameters in matter can actually be established as a by-product. To see this clearly, we first re-express the exact formulas of neutrino oscillation probabilities in Eq.~(\ref{eq:Pab_eff}) in terms of effective parameters in matter. Starting with the disappearance channel $\nu^{}_e \to \nu^{}_e$, we have
\begin{eqnarray}
\widetilde P_{ee}^{}  & = &
1 -2 c^2_{\widetilde \theta^{}_{13}} s^2_{\widetilde \theta^{}
_{13}}  \Big[1 - \cos\frac{ (\widetilde\Delta^{}_{31} +
\widetilde\Delta^{}_{32} )L}{4 E}  \cos(\frac{\widetilde\Delta^{}_{21} L}{4 E} ) \Big] \nonumber \\
&&
-2 c^{}_{2 \widetilde \theta^{}_{12}} c^2_{\widetilde \theta^{}
_{13}} s^2_{\widetilde \theta^{}_{13}}  \sin\frac{ (\widetilde
\Delta^{}_{31}
+\widetilde \Delta^{}_{32} )L}{4 E}  \sin(\frac{\widetilde\Delta^{}_{21} L}
{4 E} )  \nonumber\\ &&
- 4  c^2_{\widetilde \theta^{}_{12}} s^2_{\widetilde \theta^{}_{12}} c^4_{\widetilde \theta^{}_{13}} \sin^2 (\frac{\widetilde \Delta^{}_{21}
L}{4 E} ) \; ,
\label{eq:Pee_eff}
\end{eqnarray}
where $c^{}_{\widetilde \theta^{}_{ij}} \equiv \cos \widetilde{\theta}^{}_{ij}$ and $s^{}_{\widetilde \theta^{}_{ij}} \equiv \sin \widetilde{\theta}^{}_{ij}$ have been defined as before. Comparing between $\widetilde{P}^{}_{ee}$ in Eq.~(\ref{eq:Pee_eff}) and $\widetilde{P}^{}_{\alpha \beta}$ with $\alpha = \beta = e$ in Eq.~(\ref{eq:Pab_dis}),  one can immediately realize
\begin{eqnarray}
c^2_{\widetilde \theta^{}_{13}} s^2_{\widetilde \theta^{}_{13}} = \xi^{ee}_1 \xi^{ee}_2\; , \quad  c^{}_{2 \widetilde \theta^{}_{12}} c^2_{\widetilde \theta^{}_{13}} s^2_{\widetilde \theta^{}_{13}} = 2\xi^{ee}_1 \xi^{ee}_3/\epsilon \; , \quad c^2_{\widetilde \theta^{}_{12}} s^2_{\widetilde \theta^{}_{12}} c^4_{\widetilde \theta^{}_{13}} = (\xi^{ee}_2)^2/4 - (\xi^{ee}_3)^2/\epsilon^2 \; ,
\label{eq:map_ee}
\end{eqnarray}
by identifying the oscillation terms of the same kind. In the derivation of Eq.~(\ref{eq:map_ee}), we have implemented the following relations
\begin{eqnarray}
\frac{(\widetilde\Delta^{}_{31} + \widetilde\Delta^{}_{32}) L}
{4 E}  =   F_*^{} (3 \lambda_3^{} +b )  \;, \quad \frac{\widetilde\Delta^{}_{21}  L}
{4 E}   =   F_*^{} \epsilon  \;,
\label{eq:phase}
\end{eqnarray}
which are verified by using $\lambda^{}_i = [\widetilde{m}^2_i - (m^2_2 - \eta \Delta^{}_{21})]/\Delta^{}_*$ and $\widetilde{\Delta}^{}_{ji} \equiv \widetilde{m}^2_j - \widetilde{m}^2_i$. Thus far, the results are exact and no approximations have been made. To get more useful results for effective mixing angles $\widetilde \theta^{}_{12}$ and $\widetilde \theta^{}_{13}$, we have to expand $\xi^{ee}_i$ (for $i = 1, 2, 3$) with respect to $\alpha$ but keep $\epsilon$ unchanged as before.

Since $\widetilde{P}^{}_{ee}$ is independent of $\widetilde \theta^{}_{23}$ and the CP-violating phase $\widetilde \delta$, one should further consider the oscillations in the appearance channels. As an example, we study the oscillation probability in the appearance channel $\nu^{}_\mu \to \nu^{}_e$, namely,
\begin{eqnarray}
\widetilde P_{\mu e}^{}  & = & 2 c^2_{\widetilde \theta^{}_{13}} s^2_{\widetilde\theta^{}_{13}}  s^2_{\widetilde \theta^{}_{23}} \Big[1 - \cos\frac{(\widetilde\Delta^{}_{31} + \widetilde\Delta^{}_{32}
)L}{4 E}  \cos(\frac{\widetilde\Delta^{}_{21} L}{4 E}
) \Big] \nonumber\\ &&
+ 2 (c^{}_{2  \widetilde \theta^{}_{12}} c^2_{\widetilde \theta^{}
_{13}} s^2_{\widetilde \theta^{}_{13}} s^2_{\widetilde \theta^{}
_{23}} + 2 \widetilde{\cal J} \cot \widetilde \delta )\sin\frac{
(\widetilde \Delta^{}_{31}  + \widetilde \Delta^{}_{32} )L}{4 E}
\sin (\frac{\widetilde\Delta^{}_{21} L}{4 E} )  \nonumber\\ &&
+ 4 \Big[c^2_{\widetilde \theta^{}_{12}} s^2_{\widetilde \theta^{}
_{12}} c^2_{\widetilde \theta^{}_{13}} (c^2_{\widetilde \theta^{}
_{23}} - s^2_{\widetilde \theta^{}_{13}} s^2_{\widetilde \theta^{}
_{23}}) + c^{}_{2 \widetilde \theta^{}_{12}}
\widetilde{\cal J}  \cot \widetilde\delta \Big]\sin^2
(\frac{\widetilde
\Delta^{}_{21} L}{4 E} ) \nonumber\\
&& + 4 \widetilde{\cal J} \Big[\cos\frac{
(\widetilde\Delta^{}_{31} + \widetilde\Delta^{}_{32}
)L}{4 E}  - \cos(\frac{\widetilde\Delta^{}_{21} L}{4 E}
)  \Big] \sin^{} (\frac{\widetilde
\Delta^{}_{21} L}{4 E} ) \; ,
\label{eq:Pme_eff}
\end{eqnarray}
from which additional relations between $\{\widetilde{\theta}^{}_{23}, \widetilde{\cal J}\}$ and the parameters $\{\xi^{\alpha \beta}_i, \epsilon\}$, similar to those in Eq.~(\ref{eq:map_ee}) can be found. Using the series expansions of $\xi^{\alpha \beta}_i$ listed in Appendix~A and setting $\eta = \cos^2 \theta^{}_{12}$, we finally arrive at the mapping for three mixing angles
\begin{eqnarray}
 s^{2}_{\widetilde\theta_{13}^{}}  & \approx & \frac{s^{2}_{\theta_{13}^{}}
 (1 + \widehat A + \widehat C  )}{\widehat C  (1 - \widehat A + \widehat C  )} -  \frac{\alpha^2  (1 - \widehat A - \widehat C  )
 (1 - \widehat A^2 + 3 \widehat C - \widehat A \widehat C  )
}{4 \widehat C^3  (1 + \widehat A + \widehat C  )^2} s^{2}_{2 \theta_{12}^{}} c^{2}_{\theta_{13}^{}} \; , \nonumber \\
s^{2}_{\widetilde \theta_{12}^{}}  & \approx &
\frac{1 + 2\epsilon  + \widehat A - \widehat C
- 2 \alpha c^{}_{2\theta_{12}^{}} }{4 \epsilon} - \frac{\alpha^2
\widehat A  (2 + 3 \widehat A - 6 c^{2}_{\theta_{13}^{}} \widehat A
+ \widehat A^2 + 6 \widehat C - \widehat A \widehat C  )}
{2 \epsilon \widehat C  (1 - \widehat A + \widehat C  )^2  (1 + \widehat A + \widehat C)} s^{2}_{2\theta_{12}^{}} s^{2}_{\theta_{13}^{}} \; ,
\label{eq:map_mix}
\\
 s^{2}_{\widetilde\theta_{23}^{}}  & \approx & s^{2}_{\theta_{23}^{}} - \frac{8 \alpha {\cal J} (1 - \widehat A - \widehat C  )
 (1 + \widehat A + \widehat C + 2\alpha c^{}_{2\theta_{12}^{}}
 )\cot\delta  } {s^{2}_{2 \theta_{13}^{}}
 (1 + \widehat A + \widehat C  )^2} + \frac{\alpha^2
 (1 - \widehat A - \widehat C  )^2}{
4s^{2}_{\theta_{13}^{}}
 (1 + \widehat A + \widehat C  )^2} s^{2}_{2\theta_{12}^{}}
c^{}_{2\theta_{23}^{}} \; , \quad \nonumber
\end{eqnarray}
and that for the Jarlskog invariant
\begin{eqnarray}
\widetilde {\cal J}  & \simeq & \frac{2 \alpha {\cal J}}{\epsilon \widehat C
 (1 + \widehat A + \widehat C  )} \left[ 1 + \frac{\alpha c^{}_{2\theta_{12}^{}} (1 - \widehat{A} - \widehat{C})} {1 + \widehat A + \widehat C } \right]\; .
\label{eq:map_J}
\end{eqnarray}
Note that the leptonic CP violation is now described by the Jarlskog invariant, and the direct relation between $\widetilde{\delta}$ and the vacuum mixing parameters can be easily obtained using Eqs.~(\ref{eq:map_mix}) and (\ref{eq:map_J}). In Appendix C we also list the mapping of the three mixing angles and the Jarlskog invariant for a generic value of $\eta$.

As a cross check, we further use the relations derived in Eqs.~(\ref{eq:map_mix}) and (\ref{eq:map_J}) to calculate the following oscillation probability of $\widetilde P_{\tau \mu}^{}$
\begin{eqnarray}
\widetilde P_{\tau \mu}^{}  & = &
2 c^4_{\widetilde \theta^{}_{13}}  c^2_{\widetilde \theta^{}_{23}}
s^2_{\widetilde \theta^{}_{23}}  \Big[1 - \cos\frac{
(\widetilde\Delta^{}_{31} + \widetilde\Delta^{}_{32}
)L}{4 E}  \cos(\frac{\widetilde\Delta^{}_{21} L}{4 E}
) \Big] \nonumber\\ &&
- 2 \Big[c^{}_{2 \widetilde \theta^{}_{12}}
c^2_{\widetilde \theta^{}_{13}} c^2_{\widetilde \theta^{}_{23}}
s^2_{\widetilde \theta^{}_{23}} (1 + s^2_{\widetilde \theta^{}_{13}}) + 2
c^{}_{2 \widetilde \theta^{}_{23}} \widetilde{\cal J} \cot
\widetilde \delta \Big]\sin
\frac{ (\widetilde \Delta^{}_{31}  + \widetilde \Delta^{}_{32} )L}
{4 E}  \sin (\frac{\widetilde\Delta^{}_{21} L}{4 E} )
\nonumber\\ &&
+ 4 \Big[ c^2_{\widetilde \theta^{}_{12}}
s^2_{\widetilde\theta^{}_{12}} s^2_{\widetilde \theta^{}_{13}} +
s^2_{\widetilde \theta^{}_{13}} c^2_{\widetilde \theta^{}_{23}}
s^2_{\widetilde \theta^{}_{23}} -
c^2_{\widetilde \theta^{}_{12}} s^2_{\widetilde \theta^{}
_{12}} c^2_{\widetilde \theta^{}_{23}} s^2_{\widetilde \theta^{}
_{23}} (1 + 4 s^2_{\widetilde \theta^{}_{13}} +
s^4_{\widetilde\theta^{}_{13}}) \nonumber\\
&& + c^{}_{\widetilde\theta^{}_{12}} s^{}_{\widetilde \theta^{}
_{12}} s^{}_{\widetilde\theta^{}_{13}} c^{}_{\widetilde \theta^{}
_{23}} s^{}_{\widetilde \theta^{}_{23}}
c^{}_{2 \widetilde \theta^{}_{12}}  c^{}_{2 \widetilde \theta^{}
_{23}} (1 + s^{2}_{\widetilde \theta^{}_{13}}) \cos\widetilde
\delta - 2 c^{2}_{\widetilde \theta^{}_{12}}
s^{2}_{\widetilde\theta^{}_{12}} s^{2}_{\widetilde \theta^{}_{13}}
c^{2}_{\widetilde \theta^{}_{23}} s^{2}_{\widetilde \theta^{}_{23}} \cos 2 \widetilde \delta \Big]\sin^2
(\frac{\widetilde\Delta^{}_{21} L}{4 E} ) \nonumber\\
&& + 4 \widetilde{\cal J} \Big[\cos\frac{
(\widetilde\Delta^{}_{31} + \widetilde\Delta^{}_{32}
)L}{4 E}  - \cos(\frac{\widetilde\Delta^{}_{21} L}{4 E}
)  \Big] \sin^{} (\frac{\widetilde
\Delta^{}_{21} L}{4 E} )\; .
\label{eq:Ptm_eff}
\end{eqnarray}
It turns out that the expression in Eq.~(57) can be exactly reproduced, when both of them are matched to the same order of $\alpha$.

\begin{figure}[H]
\centerline{
\hspace{0cm}
\includegraphics[scale=0.96]{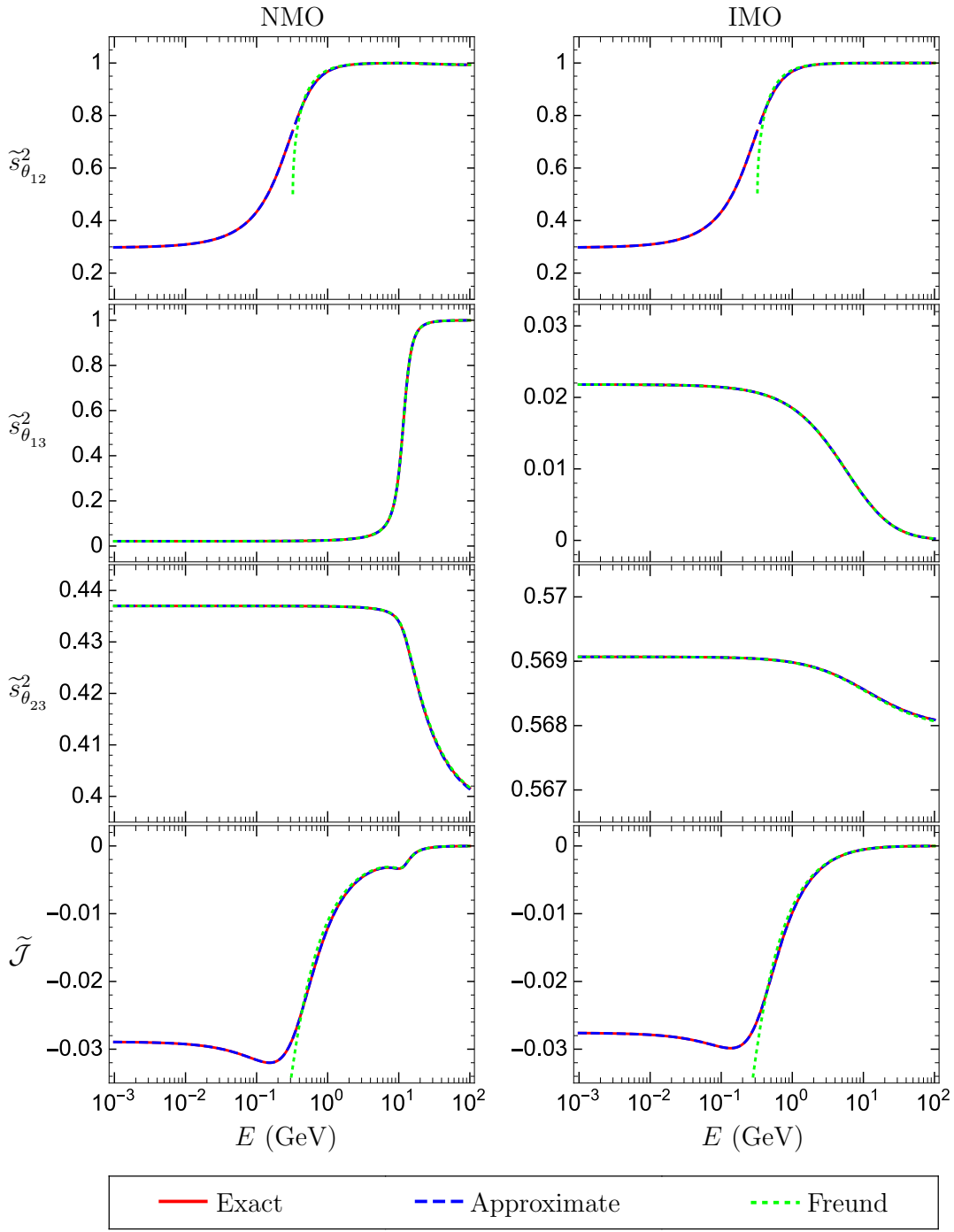}
}
\caption{Three effective mixing angles $\{\sin^2 \widetilde\theta^{}_{12}, \sin^2 \widetilde\theta^{}_{13}, \sin^2 \widetilde\theta^{}_{23}\}$ and the effective Jarlskog invariant $\widetilde{\cal J}$ shown as functions of neutrino energies, where $\eta = \cos^2 \theta^{}_{12}$, a constant matter density of $\rho \approx 2.8~{\rm g}~{\rm cm}^{-3}$ with the electron fraction $Y^{}_e \approx 0.5$, and the best-fit values of neutrino oscillation parameters in Table~1 have been used.}
\label{fig:map}
\end{figure}

It is worth mentioning that the mapping relations for mixing angles and the Jarlskog invariant have been truncated at the second order of $\alpha$ and serve as excellent approximations to the exact results. For illustration, we have calculated the effective mixing angles $\{\sin^2 \widetilde\theta^{}_{12}, \sin^2 \widetilde\theta^{}_{13}, \sin^2 \widetilde\theta^{}_{23}\}$ and the effective Jarlskog invariant $\widetilde{\cal J}$ for different neutrino energies. As depicted in Fig.~\ref{fig:map}, the exact results are denoted as solid curves (red), while the approximate results based on Eqs.~(\ref{eq:map_mix}) and (\ref{eq:map_J}) are represented by dashed curves (blue). One can see that our approximate results are in perfect agreement with the exact ones, and the differences between them are invisible from the plots. For comparison, the numerical results according to the mapping relations found by Freund in Ref.~\cite{Freund:2001pn} are given as dotted curves (green). Significant deviations can be observed in the figures for $\sin^2 \widetilde{\theta}^{}_{12}$ and $\widetilde{J}$, which can be explained by the divergence encountered in the low-energy region.
\begin{figure}[!t]
\centerline{
\hspace{0cm}
\includegraphics[scale=0.96]{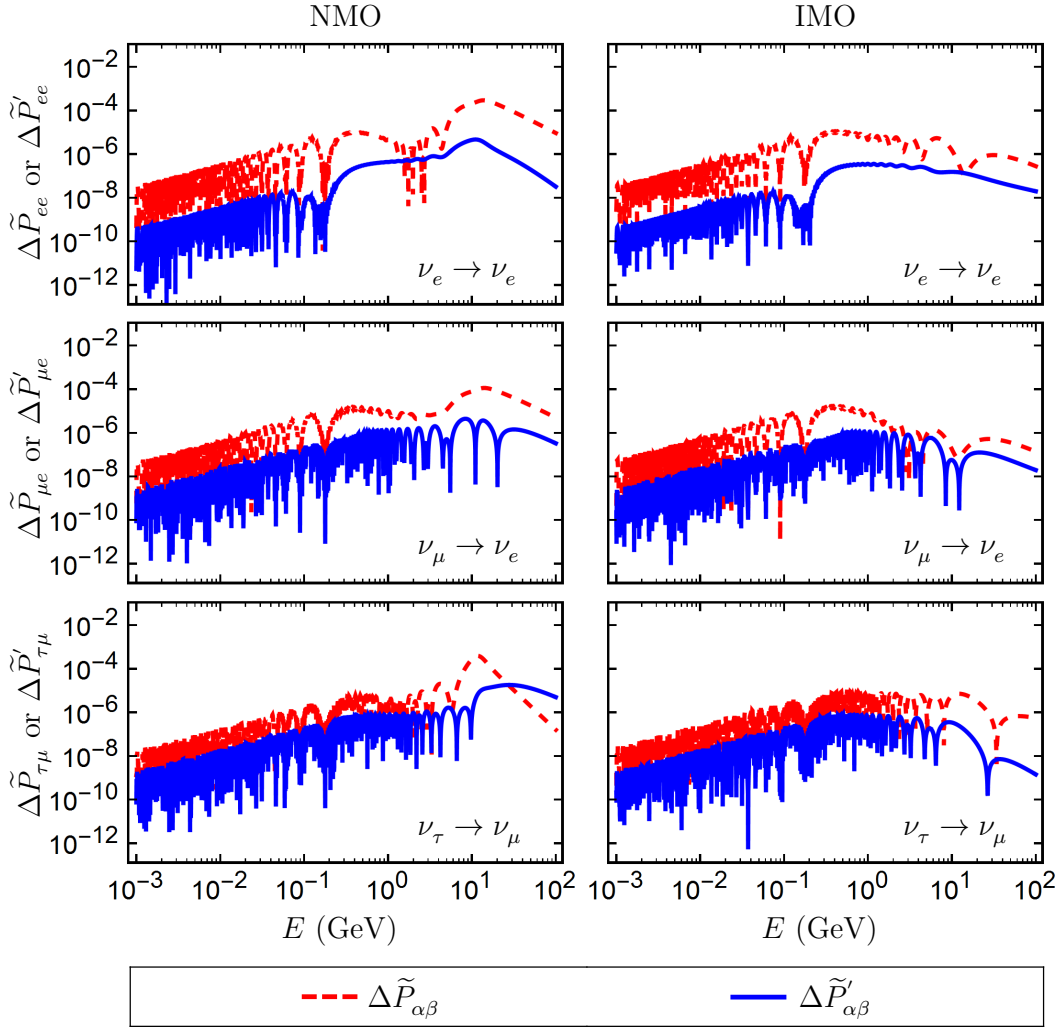}
}
\caption{Numerical comparison between $\Delta \widetilde P_{\alpha\beta}^{'}$
and $\Delta \widetilde P_{\alpha\beta}^{}$, where $\eta = \cos^2 \theta^{}_{12}$ is fixed and the other input parameters are the same as in Fig.~\ref{fig:error}.}
\label{fig:newP}
\end{figure}

Moreover, given the approximate expressions of effective mixing parameters in Eqs.~(\ref{eq:map_mix}) and (\ref{eq:map_J}), one can insert them back into Eqs.~(\ref{eq:Pee_eff}), (\ref{eq:Pme_eff}) and (\ref{eq:Ptm_eff}) and obtain a new set of oscillation probabilities, which we call $\widetilde P_{ee}^{'}$, $\widetilde P_{\mu e}^{'}$ and $\widetilde P_{\tau \mu}^{'}$, respectively.  As the effective mixing parameters are expanded up to ${\cal O}(\alpha^2)$, these new oscillation probabilities will be more accurate in the sense that part of higher-order terms are now included. To illustrate this point, we compute the absolute errors $\Delta \widetilde P_{\alpha \beta}^{'}$ according to Eq.~(\ref{eq:DeltaP}) and compare it with $\Delta \widetilde{P}^{}_{\alpha \beta}$ from Fig.~\ref{fig:error} in the case of $\eta = \cos^2 \theta^{}_{12}$. The results are shown in Fig.~\ref{fig:newP}, where one can find $\Delta \widetilde P_{\alpha \beta}^{'}$ (blue solid curves) are almost always one or two orders of magnitude smaller than $\Delta \widetilde{P}^{}_{\alpha \beta}$ (red dashed curves).

\section{Summary}

In this work we have taken a deep look into analytical approximations for three-flavor neutrino oscillation probabilities in matter of a constant density and presented a new set of simple and compact formulas. A useful definition of the $\eta$-gauge neutrino mass-squared difference $\Delta^{}_* \equiv \eta \Delta^{}_{31} + (1-\eta) \Delta^{}_{32}$
is introduced and the calculations are performed in the series expansions of $\alpha$ (i.e., $\alpha\equiv\Delta^{}_{21}/\Delta^{}_*$).
The approximate oscillation probabilities are valid for arbitrary neutrino energies and any baseline length. Among different
choices of $\eta$, it turns out that the case of $\eta = \cos^2\theta^{}_{12}$ is the best one in terms of both simplicity and numerical accuracy. These formulas are particularly useful for the future long baseline accelerator neutrino experiments and the atmospheric neutrino experiments with the baseline lengths from $10~{\rm km}$ to $ 10^4~{\rm km}$ and a wide range of neutrino energies ($0.1~{\rm GeV} \lesssim E \lesssim 100~{\rm GeV}$). The main features of our results can be summarized as follows.
\begin{itemize}
\item Our calculations are based on the Cayley-Hamilton theorem, where only the effective Hamiltonian and its three eigenvalues are needed in order to derive the oscillation probabilities. The series expansions of $\alpha$ are applied to the exact expressions of the eigenvalues in Eq.~(\ref{eq:lambda_NMO}) or Eq.~(\ref{eq:lambda_IMO}). However, the $\epsilon$ parameter in the expansions of $\lambda_1$ and $\lambda_2$ in Eq.~(\ref{eq:lambda_app}) behaves as the function $\widehat{\epsilon} \equiv (\alpha^2 + \widehat{A}^2 \cos^4\theta^{}_{13} - 2\widehat{A}\alpha \cos2\theta^{}_{12} \cos^2\theta^{}_{13})^{1/2}$ and cannot be expanded in terms of $\alpha$ in the low energy range with $\widehat{A} \lesssim \alpha$. Thus, we keep $\epsilon$ intact in the calculations.

\item Our calculations employ a generic $\eta$-gauge neutrino mass-squared difference $\Delta^{}_*$ and derive the $\eta$-gauge oscillation probabilities as shown in Eqs.~(\ref{eq:Pee}), (\ref{eq:Pme}) and (\ref{eq:Ptm}) for $\widetilde P_{ee}^{}$, $\widetilde P_{\mu e}^{}$ and $\widetilde P_{\tau\mu}^{}$ respectively. Given the expressions of $\rho_{i}$ ($i=1,2,3$) in Eq.~(\ref{eq:rhoi}) and $\epsilon^2$ in Eq.~(\ref{eq:epsilon}), the analytical results of oscillation probabilities are greatly simplified for $\eta = \cos^2 \theta^{}_{12}$, where all the terms proportional to
$(\eta-\cos^2 \theta^{}_{12})$ automatically disappear. Moreover, as demonstrated in Fig.~\ref{fig:error} for different values of $\eta$,
the choice of $\eta = \cos^2 \theta^{}_{12}$ is the most accurate one for almost the entire range of neutrino energies.

\item Fixing the gauge at $\eta = \cos^2 \theta^{}_{12}$, the oscillation probabilities are presented in Eqs.~(\ref{eq:Pee_new}), (\ref{eq:Pme_new}) and (\ref{eq:Ptm_new}) for $\widetilde P_{ee}^{}$, $\widetilde P_{\mu e}^{}$ and $\widetilde P_{\tau\mu}^{}$ respectively, constituting the main results of this work. Regarding the accuracy of these analytical approximations, a careful study is performed in Fig.~\ref{fig:2D} for the neutrino energies from $10^{-3}$ GeV to $10^{2}$ GeV and
the baseline length range $10^{-1}~{\rm km} \leq L \leq 10^{4}~{\rm km}$. One can observe that in the NMO case the errors in the entire range of baseline lengths and neutrino energies are below $10^{-3}$, while for IMO below $10^{-4}$. The largest errors appear in NMO around $E \approx 10~{\rm GeV}$ and $L \approx 5000~{\rm km}$, where the atmospheric resonance is encountered and the small energy splitting between $\lambda_2$ and $\lambda_3$ slows down the convergence of the series expansions.

\item As a by-product a more accurate mapping of the intrinsic mixing parameters to the effective mixing parameters in matter is established in Eqs.~(\ref{eq:map_mix}) and (\ref{eq:map_J}) for three mixing angles and the Jarlskog invariant, respectively. With the effective mixing parameters, one can obtain a new set of oscillation probabilities in Eqs.~(\ref{eq:Pee_eff}), (\ref{eq:Pme_eff}) and (\ref{eq:Ptm_eff}) for $\widetilde P_{ee}^{'}$, $\widetilde P_{\mu e}^{'}$ and $\widetilde P_{\tau \mu}^{'}$, respectively. The accuracy of the effective mixing parameters is proved in Fig.~\ref{fig:map} for the whole energy range including the regions of the solar and atmospheric resonances. For the new set of oscillation probabilities, one can find from Fig.~\ref{fig:newP} that the accuracy of $\widetilde P_{\alpha \beta}^{'}$ will be one or two orders of magnitude better than $\widetilde{P}^{}_{\alpha \beta}$ because some higher-order terms are also properly included.

\item Finally, in the low energy range, $\alpha$, $\widehat{A}$ and $\epsilon$ are of the same order and can be expanded simultaneously, from which one can arrive at Eq.~(4.6) in Ref.~\cite{Xing:2016ymg}. On the other hand, for the high energy region with $\widehat{A}\sim\epsilon\gg\alpha$, one can safely
expand $\epsilon$ in terms of $\alpha$ and restore the familiar results of Freund~\cite{Freund:2001pn} and Akhmedov {\it et al.}~\cite{Akhmedov:2004ny}
\end{itemize}

For future long-baseline accelerator and atmospheric neutrino experiments, with the goals of determining the neutrino mass ordering and measuring the leptonic CP violating phase, a set of compact and simple analytical approximations of oscillation probabilities in matter is very helpful. These analytical oscillation probabilities should be directly connected with the fundamental oscillation parameters and be valid for arbitrary neutrino energies and any baseline length.
In this sense, our analytical approximations in this work meet all the afore-mentioned criteria and can be readily applied to future oscillation experiments. We leave such applications for a separate work in the near future.

\appendix

\section{Expressions for the $\xi_{i}^{ \alpha\beta}$ terms}

In this appendix, we present expressions for the $\xi_{i}^{ \alpha\beta}$ terms, which are coefficients in front of various oscillation terms in Eq.~(\ref{eq:Pab_s}). For this purpose, we employ the Cayley-Hamilton theorem to express the evolution matrix $S = e^{-2\mathrm{i} F_*^{} M_{\rm f}^{}}_{}$ into a form similar to Eq.~(\ref{eq:CH}) with $\widetilde{H}_{\rm f}^{}$ replaced by $M_{\rm f}^{}$. Correspondingly, $\omega_i^{}$ in Eq.~(\ref{eq:si}) are now the eigenvalues of $M_{\rm f}^{}$, i.e., $\lambda_i^{}$, and $e^{-\mathrm{i}\omega_i^{} L}$ read as $e^{-2 \mathrm{i} F_*^{} \lambda_i^{}}$.  Then, with the explicit form of $M_{\rm f}^{}$ given in Eq.~(\ref{eq:M_f}), we are able to obtain the expressions of various $\xi_i^{\alpha\beta}$ for various oscillation channels $\widetilde{P}_{\alpha\beta}$, according to the definitions of $\xi_i^{\alpha\beta}$ in Eq.~(\ref{eq:Pab_s}).

As shown in Eq.~(\ref{eq:Pab_full}), the final oscillation probabilities $\widetilde{P}_{\alpha\beta}$ only depend on certain combinations of $\xi_i^{\alpha\beta}$,  we therefore just show the analytical expansions for those relevant ones. For $\widetilde P_{ee}^{}$, we have $\xi_{1}^{ee} + \xi_{2}^{ee} =1$, $\xi_{i}^{ee} = \xi_{i}^{ee*}$ and
\begin{eqnarray}
\xi_1^{ee} \xi_2^{ee} &\approx &
\frac{s^{2}_{2 \theta_{13}^{}} }{4 \widehat C^2} - \frac{\alpha\widehat A
 ( \eta -c^{2}_{\theta_{12}^{}}  )  (
\widehat A - c^{}_{2\theta_{13}^{}} )}
{2 \widehat C^4} s^{2}_{2 \theta_{13}^{}} \;,
\label{eq:xi_ee1}
\end{eqnarray}
\begin{eqnarray}
\xi_1^{ee} \xi_3^{ee}  &\approx &
\frac{-\widehat A c^{4}_{\theta_{13}^{}} s^{2}_{\theta_{13}^{} }
}{\widehat C^2  ( 1 + \widehat A + \widehat C  ) } + \frac{ \alpha
c^{}_{2 \theta_{12}^{}} s^{2}_{2\theta_{13}^{}} }{8 \widehat C^2}
+ \frac{\alpha \widehat A  ( \eta - c^{2}_{\theta_{12}^{}}  )
}{2 \widehat C^4  ( 1 + \widehat A + \widehat C  )}
\nonumber\\ &&
\times(1 - 6 \widehat A c^{}_{2\theta_{13}^{}} - \widehat C + \widehat A \widehat C + 5 \widehat A^2)
c^{4}_{\theta_{13}^{}} s^{2}_{\theta_{13}^{}} \;,
\label{eq:xi_ee2}
\end{eqnarray}
\begin{eqnarray}
\left(\xi_3^{ee}\right)^2_{} - \frac{\epsilon^2}{4} \left(\xi_2^{ee}\right)^2_{} &\approx &
-\frac{\alpha^2  ( 1 - \widehat A + \widehat C  )}
{\widehat C  ( 1 + \widehat A + \widehat C  )^3} s^{2}_{2\theta_{12}^{}}
c^{4}_{\theta_{13}^{}} \;.
\label{eq:xi_ee3}
\end{eqnarray}
For $\widetilde P_{\mu e}^{}$, we have $\xi_{1}^{\mu e} = -\xi_{2}^{\mu e}$ and
\begin{eqnarray}
\xi^{\mu e}_1 \xi^{\mu e*}_1  & \approx &
\frac{ s^2_{2\theta_{13}^{}} s^2_{\theta_{23}^{}} }{4 \widehat C^2} -
\frac{ 2 \alpha  ( 1 - \widehat A - \widehat C) }{\widehat C^2  ( 1 + \widehat A
+ \widehat C  ) } {\cal J} \cot \delta
\nonumber\\ &&
- \frac{\alpha\widehat A
 ( \eta -c^2_{\theta_{12}^{}} )  ( \widehat A -
c^{}_{2 \theta_{13}^{}}  )}{2 \widehat C^4}
s^2_{2\theta_{13}^{}} s^2_{\theta_{23}^{}} ,
\label{eq:xi_me1}
\end{eqnarray}
\begin{eqnarray}
\xi^{\mu e}_1 \xi^{\mu e*}_3 + \xi^{\mu e*}_1 \xi^{\mu e}_3
& \approx &  \frac{  ( 1 +
\widehat A - \widehat C)} {8 \widehat C^2 } s^2_{2 \theta_{13}
^{}} s^2_{\theta_{23}^{}}
- \frac{\alpha  ( 1 - \widehat A + \widehat C
 )}{\widehat C^2} {\cal J} \cot \delta
 \nonumber\\ &&
 -\frac{\alpha }{4 \widehat C^2}
c^{}_{2\theta_{12}^{}} s^2_{2 \theta_{13}^{}} s^2_{\theta_{23}
^{}} - \frac{\alpha \widehat A  ( \eta -
c^2_{\theta_{12}^{}} )
}{\widehat C^4  ( 1 + \widehat A + \widehat C) }   (
1 - 6 \widehat A c^{}_{2\theta_{13}^{}}
\nonumber \\ &&
 - \widehat C + \widehat A \widehat C + 5 \widehat A^2
 ) c^4_{\theta_{13}^{}} s^2_{\theta_{13}^{}}
s^2_{\theta_{23}^{}} ,
\label{eq:xi_me2}
\end{eqnarray}
\begin{eqnarray}
\xi^{\mu e}_1 \xi^{\mu e*}_3 - \xi^{\mu e*}_1 \xi^{\mu e}_3
&\approx & \frac{4 i \alpha {\cal J}}
{\widehat C  ( 1 + \widehat A + \widehat C)}  ,
\label{eq:xi_me3}
\end{eqnarray}
\begin{eqnarray}
\xi^{\mu e}_3 \xi^{\mu e*}_3 - \frac{\epsilon^2}{4} \xi^{\mu e}_1 \xi^{\mu e*}_1
  & \approx &
\frac{\alpha^2  (1 - \widehat A + \widehat C)}{
4 \widehat C  (1 + \widehat A + \widehat C  )}
s^{2}_{2\theta_{12}^{}} c^{2}_{\theta_{13}^{}}
c^{2}_{\theta_{23}^{}}
 \nonumber\\ &&
- \frac{\alpha^2
(1 + \widehat A)}{4 \widehat C(1 + \widehat A + \widehat C)^2}
s^{2}_{2\theta_{12}^{}} s^{2}_{2 \theta_{13}^{}}
s^{2}_{\theta_{23}^{}}
\nonumber \\ & &
\displaystyle+ \frac{4 \alpha {\cal J \cot \delta}}{\widehat C
 (1 + \widehat A + \widehat C  )^2}  \Bigr[ \alpha c^{}_{ 2\theta_{12}
^{}}  (\widehat C + \widehat A c^2 _{\theta_{13}^{}} ) - \widehat A
c^2 _{\theta_{13}^{}} \Bigr]
\nonumber \\ & & +
\frac{4 \alpha^2 \widehat A  ( \eta -c^2_{\theta_{12}^{}} )}
{\widehat C^3  ( 1 + \widehat A + \widehat C)^2}  ( 1 - 3 \widehat A c^{}
_{2\theta_{13}^{}}
 \nonumber\\ & &
- \widehat C + \widehat A \widehat C + 2 \widehat A^2  )
c^2_{\theta_{13}^{}} {\cal J } \cot\delta \;.
\label{eq:xi_me4}
\end{eqnarray}
For $\widetilde P_{\tau \mu}^{}$, we have $\xi_{1}^{\tau \mu} = -\xi_{2}^{\tau \mu}$ and
\begin{eqnarray}
\xi^{\tau \mu}_1 \xi^{\tau \mu*}_1  & \approx &
\frac{  (1 - \widehat A + \widehat C  ) ^2}
{4 \widehat C^2  (1 + \widehat A + \widehat C  )^2} c^{4}_{\theta_{13}^{}}
s^{2}_{2\theta_{23}^{}}
\nonumber\\ & &
+ \frac{8 \alpha \widehat A  (1 - \widehat A + \widehat C  )
}{\widehat C^2  (1 + \widehat A + \widehat C  )^3} c^{2}_{\theta_{13}^{}}
 c^{}_{2 \theta_{23}^{}} {\cal J} \cot\delta
\nonumber\\ & & -
\frac{\alpha \widehat A  ( \eta -c^2_{\theta_{12}^{}} )
 ( 1 - \widehat A + \widehat C)}{\widehat C^4  ( 1 + \widehat A + \widehat C)}
c^4_{\theta_{13}^{}} s^2_{\theta_{13}^{}} s^2_{2 \theta_{23}^{}}
\; ,
\label{eq:xi_tm1}
\end{eqnarray}
\begin{eqnarray}
\xi^{\tau \mu}_1 \xi^{\tau \mu*}_3 + \xi^{\tau \mu*}_1 \xi^{\tau \mu}_3
& \approx &
\frac{  (1 - \widehat A + \widehat C  )
}{16 \widehat C^2  (1 + \widehat A + \widehat C  )}
(1 + \widehat A - \widehat C -2 \alpha c^{}_{2\theta_{12}^{}} )
\nonumber \\ &&\times
(c^{}_{2\theta_{13}^{}} - \widehat A - 3 \widehat C)
c^{2}_{\theta_{13}^{}} s^{2}_{2\theta_{23}^{}}
+ \alpha c^{}_{2\theta_{23}^{}} {\cal J}\cot\delta
\nonumber \\ &&\times
\Bigr [\frac{1 + \widehat A + \widehat C}{ \widehat C^2} -
\frac{8 \widehat A (\widehat A c^{2}_{\theta_{13}^{}} + \widehat C
+ \widehat A \widehat C)}{ \widehat C^2  (1 + \widehat A + \widehat C  )^2} \Bigr]
\nonumber\\ &&
+ \frac{\alpha \widehat A^2  ( \eta -c^2_{
\theta_{12}^{}} )  }{2 \widehat C^4  ( 1 + \widehat A + \widehat C)^2}
\Bigr[ ( 1 + \widehat A  )  (7 + 7\widehat A + 5 \widehat C  )
\nonumber\\ &&
-2 c^2_{\theta_{13}^{}} (
2 + 12\widehat A + 3 \widehat C) \Bigr] c^4_{\theta_{13}^{}}
s^2_{\theta_{13}^{}} s^2_{2\theta_{23}^{}}  \;,
\label{eq:xi_tm2}
\end{eqnarray}
\begin{eqnarray}
\xi^{\tau \mu}_1 \xi^{\tau \mu*}_3 - \xi^{\tau \mu*}_1 \xi^{\tau \mu}_3
& \approx &  \frac{4 i \alpha {\cal J}}
{\widehat C  ( 1 + \widehat A + \widehat C)}  \;,
\label{eq:xi_tm3}
\end{eqnarray}
\begin{eqnarray}
\xi^{\tau \mu}_3 \xi^{\tau \mu*}_3 - \frac{\epsilon^2}{4} \xi^{\tau \mu}_1 \xi^{\tau \mu*}_1
& \approx &
-\frac{\widehat A  (1 - \widehat A - \widehat C)}{4 \widehat C  (1 +
\widehat A + \widehat C  )} c^{4}_{\theta_{13}^{}} s^{2}_{2\theta_{23}^{}}
+ \frac{\alpha  (1 - \widehat A - \widehat C  )}{4 \widehat C}
c^{}_{2\theta_{12}^{}} c^{2}_{\theta_{13}^{}}
s^{2}_{2\theta_{23}^{}} \nonumber\\
&& + \frac{\alpha \widehat A  ( \eta -c^2_{\theta_{12}^{}} )
 ( 1 - \widehat A - \widehat C)}{4 \widehat C^3  ( 1 + \widehat A + \widehat C)}
 ( 1 - 3 \widehat A c^{}
_{2\theta_{13}^{}} - \widehat C  + 2 \widehat A^2  )
\nonumber\\
&& \times c^4_{\theta_{13}^{}} s^2_{2 \theta_{23}^{}}
+ \frac{\alpha  (1 - \widehat A - \widehat C-
2\widehat A^2 - 2\widehat A \widehat C  )}{\widehat C  (1 + \widehat A + \widehat C  )}
{\cal J} c^{}_{2\theta_{23}^{}} \cot\delta
\nonumber \\
&& +  \alpha^2 c^{}_{2\theta_{12}^{}}  c^{}
_{2\theta_{23}^{}} {\cal J} \cot\delta
 \Bigr[\frac{4}{ c^{2}_{\theta_{13}^{}}
 (1 - \widehat A + \widehat C  )} - \frac{2}{\widehat C  (1 + \widehat A + \widehat C
 )}  \nonumber \\
&&   -\frac{ (1 - \widehat A - \widehat C  )^2
 (1 + 2 c^{2}_{\theta_{13}^{}} + \widehat A + 3 \widehat C)
}{2 \widehat C s^{2}_{\theta_{13}^{}}  (1 + \widehat A + \widehat C  )
^2 } \Bigr]
\nonumber\\ &&
- \frac{\alpha^2( 1 + \widehat A  )}{8 \widehat C}
s^{2}_{2 \theta_{12}^{}} s^{2}_{\theta_{13}^{}}
s^{2}_{2 \theta_{23}^{}} \cos 2 \delta \nonumber \\
&& + \frac{\alpha^2  ( \eta -c^2_{\theta_{12}^{}} )}
{\widehat C^3  ( 1 + \widehat A + \widehat C) }  \Bigr[2  ( \widehat C - 1)
- 5 \widehat A -2 \widehat A^2 + 3 \widehat A \widehat C
\nonumber\\ &&
+\widehat A^2  (\widehat A + \widehat C)(3 + 2 \widehat A)
-2 \widehat A c^2_{\theta_{13}^{}} (
-5 + \widehat A + 2 \widehat A^2
\nonumber\\&&
+3 \widehat C +2 \widehat A \widehat C)
 \Bigr] c^{}_{2\theta_{23}^{}} {\cal J} \cot\delta
\nonumber\\&&
+ \frac{\alpha^2  (1 + \widehat A)}{4 \widehat C} s^{2}_{2 \theta_{12}
^{}} s^{2}_{\theta_{13}^{}}
+ \frac{\alpha^2  (1 + \widehat A + \widehat C  )}{4 \widehat C  (1 -
\widehat A + \widehat C  )}s^{2}_{\theta_{13}^{}} s^{2}_{2 \theta_{23}^{}}
\nonumber \\&&
- \frac14 \alpha^2 s^{2}_{2\theta_{12}^{}} s^{2}_{2\theta_{23}
^{}}  (T)- \frac{\alpha^2  ( \eta -c^2_{\theta_{12}^{}} )}
{8 \widehat C^3}  ( 1 - \widehat A - \widehat C)
\nonumber \\&& \times
(1 - \widehat C - 4 \widehat A c^{}_{2\theta_{13}^{}}
+ 3 \widehat A^2 - \widehat A\widehat C)  c^{}_{2\theta_{12}^{}}
c^{2}_{\theta_{13}^{}} s^{2}_{2\theta_{23}^{}}
\nonumber\\&&
 + \frac{\alpha^2 \widehat A  ( \eta -c^2_{\theta_{12}^{}}
 )^2}{4 \widehat C^5  ( 1 + \widehat A + \widehat C)} \Bigm\{
- ( 1 + \widehat A  )^2  \Bigr[ 1 + 2 \widehat A + \widehat A^2
 (3 - \widehat A)\Bigr ]
 \nonumber\\ &&
 + \widehat C + \widehat A \widehat C  (
3 + \widehat A + \widehat A^3  )
 -2 \widehat A^2 c^4_{\theta_{13}^{}}  (13 -3 \widehat A
- 3 \widehat C  )
 \nonumber\\&&
-\widehat A c^2_{\theta_{13}^{}}  \Bigr[-9
+7 \widehat C -24 \widehat A +\widehat A \widehat C
\nonumber \\ &&
+ \widehat A^2  ( -11 + 4 \widehat A  + 4 \widehat C) \Bigr]
\Bigm\}
c^4_{\theta_{13}^{}} s^2_{2 \theta_{23}^{}} \;,
\label{eq:xi_tm4}
\end{eqnarray}
where
\begin{eqnarray}
T & = & \frac{1 - s^{4}_{\theta_{13}^{}}}{\widehat C(1 + \widehat A + \widehat C)^2 }
\Bigr[ 1+ \widehat A -\widehat C -2 \widehat A c^{2}_{\theta_{13}^{}}
(2 + \widehat A  ) + \widehat A^2 (1 + \widehat A + \widehat C) \Bigr] +\frac{1+\widehat A}{4 \widehat C}
(1 + 4 s^{2}_{\theta_{13}^{}} + s^{4}_{\theta_{13}^{}}) \nonumber \\
& & + \frac{c^{4}_{\theta_{13}^{}}}{\widehat C^3  (1 + \widehat A + \widehat C
 )^3}\Bigm\{-\widehat A c^{2}_{\theta_{13}^{}}
 (3-\widehat A  ) (1-\widehat A)^3  (1 - \widehat A - \widehat C  )
\nonumber\\ &&
 +\widehat A^2 s^{2}_{2\theta_{13}^{}}  (3 - 3\widehat C + 6\widehat A -3 \widehat A^2 -\widehat A \widehat C)
\nonumber\\ &&
- \widehat A s^{2}_{\theta_{13}^{}} (1+\widehat A  )\Bigr[6  (1 + \widehat A - \widehat C  ) + \widehat A^2
 (2 - \widehat A  )  (1 + \widehat A +\widehat C) \Bigr]\Bigm\}\;.
\label{eq:T}
\end{eqnarray}

\section{Expressions for $\widetilde P_{\tau\mu}^{}$}
In this appendix, we show the expression for the oscillation
probability $\widetilde P_{\tau\mu}^{}$ with an arbitrary value of $\eta$
for completeness:
\begin{eqnarray}
\widetilde P_{\tau\mu}^{}  & \approx &
\Bigr[ \frac{(1 - \widehat A + \widehat C ) ^2}
{2 \widehat C^2 (1 + \widehat A + \widehat C)^2} c^{4}_{\theta_{13}
^{}}s^{2}_{2\theta_{23}^{}}
+ \frac{16 \alpha \widehat A (1 - \widehat A + \widehat C )
}{\widehat C^2 (1 + \widehat A + \widehat C )^3} c^{2}_{\theta_{13}^{}}
 c^{}_{2 \theta_{23}^{}} {\cal J} \cot\delta \nonumber\\
&& - \frac{2 \alpha \widehat A ( \eta -c^2_{\theta_{12}^{}})
( 1 - \widehat A + \widehat C)}{\widehat C^4 ( 1 + \widehat A + \widehat C)}
c^4_{\theta_{13}^{}} s^2_{\theta_{13}^{}} s^2_{2 \theta_{23}^{}}
\Bigr](1 - \cos \widetilde F_+^{} \cos
\widetilde F_-^{})
\nonumber\\&& - \frac{2}{\epsilon}\biggr\{\frac{ (1 - \widehat A +
\widehat C )}{16 \widehat C^2 (1 + \widehat A + \widehat C )}
(1 + \widehat A - \widehat C -
2 \alpha c^{}_{2\theta_{12}^{}})
(c^{}_{2\theta_{13}^{}} - \widehat A - 3 \widehat C)
c^{2}_{\theta_{13}^{}} s^{2}_{2\theta_{23}^{}} \nonumber\\ && + \alpha c^{}_{2\theta_{23}^{}} {\cal J}\cot\delta
\Bigr[\frac{1 + \widehat A + \widehat C}{ \widehat C^2} -
\frac{8 \widehat A (\widehat A c^{2}_{\theta_{13}^{}} + \widehat C
+ \widehat A \widehat C)}{ \widehat C^2 (1 + \widehat A + \widehat C )^2}
\Bigr]  + \frac{\alpha \widehat A^2 ( \eta -c^2_{
\theta_{12}^{}})  }{2 \widehat C^4 ( 1 + \widehat A + \widehat C)^2}
\nonumber\\ && \times
\Bigr[( 1 + \widehat A ) (7 + 7\widehat A + 5 \widehat C)-2 c^2_{\theta_{13}
^{}}( 2 + 12\widehat A + 3 \widehat C) \Bigr] c^4_{\theta_{13}^{}}
s^2_{\theta_{13}^{}} s^2_{2\theta_{23}^{}}
\biggr\}\sin\widetilde F_+^{} \sin\widetilde
F_-^{} \nonumber\\ && + \biggr\{-\frac{\widehat A (1 - \widehat A - \widehat
C)}{\widehat C (1 + \widehat A + \widehat C )} c^{4}_{\theta_{13}^{}} s^{2}
_{2\theta_{23}^{}} + \frac{\alpha (1 - \widehat A - \widehat C )}{\widehat C}
c^{}_{2\theta_{12}^{}} c^{2}_{\theta_{13}^{}}
s^{2}_{2\theta_{23}^{}} \nonumber\\
&& + \frac{\alpha \widehat A ( \eta -c^2_{\theta_{12}^{}})
( 1 - \widehat A - \widehat C)}{\widehat C^3 ( 1 + \widehat A + \widehat C)}
( 1 - 3 \widehat A c^{}
_{2\theta_{13}^{}} - \widehat C  + 2 \widehat A^2 )
c^4_{\theta_{13}^{}} s^2_{2 \theta_{23}^{}}
\nonumber\\
&& + \frac{4 \alpha (1 - \widehat A - \widehat C-
2\widehat A^2 - 2\widehat A \widehat C )}{\widehat C (1 + \widehat A + \widehat C )}
{\cal J} c^{}_{2\theta_{23}^{}} \cot\delta
\nonumber \\
&& +  4 \alpha^2 c^{}_{2\theta_{12}^{}}  c^{}
_{2\theta_{23}^{}} {\cal J} \cot\delta
\Bigr[\frac{4}{ c^{2}_{\theta_{13}^{}}
(1 - \widehat A + \widehat C )} - \frac{2}{\widehat C (1 + \widehat A + \widehat C
)} .\nonumber \\
&& -\frac{(1 - \widehat A - \widehat C )^2
(1 + 2 c^{2}_{\theta_{13}^{}} + \widehat A + 3 \widehat C)
}{2 \widehat C s^{2}_{\theta_{13}^{}} (1 + \widehat A + \widehat C )
^2 }\Bigr] - \frac{\alpha^2( 1 + \widehat A )}{2 \widehat C}
s^{2}_{2 \theta_{12}^{}} s^{2}_{\theta_{13}^{}}
s^{2}_{2 \theta_{23}^{}} \cos 2 \delta \nonumber \\
&& + \frac{4 \alpha^2 ( \eta -c^2_{\theta_{12}^{}})}
{\widehat C^3 ( 1 + \widehat A + \widehat C) } \Bigr[2 ( \widehat C - 1)
- 5 \widehat A -2 \widehat A^2 + 3 \widehat A \widehat C +\widehat A^2 (\widehat A + \widehat
C)(3 + 2 \widehat A) \nonumber\\
&&-2 \widehat A c^2_{\theta_{13}^{}} (
-5 + \widehat A + 2 \widehat A^2 +3 \widehat C +2 \widehat A \widehat C)
\Bigr] c^{}_{2\theta_{23}^{}} {\cal J} \cot\delta \nonumber\\
&&+ \frac{\alpha^2 (1 + \widehat A)}{ \widehat C} s^{2}_{2 \theta_{12}
^{}} s^{2}_{\theta_{13}^{}}
+ \frac{\alpha^2 (1 + \widehat A + \widehat C )}{\widehat C (1 -
\widehat A + \widehat C )}s^{2}_{\theta_{13}^{}} s^{2}_{2 \theta_{23}^{}}
-  \alpha^2 s^{2}_{2\theta_{12}^{}} s^{2}_{2\theta_{23}
^{}} (T)\nonumber \\
&&- \frac{\alpha^2 ( \eta -c^2_{\theta_{12}^{}})}
{2 \widehat C^3} ( 1 - \widehat A - \widehat C) (
1 - \widehat C - 4 \widehat A c^{}_{2\theta_{13}^{}}
+ 3 \widehat A^2 - \widehat A\widehat C)  c^{}_{2\theta_{12}^{}}
c^{2}_{\theta_{13}^{}} s^{2}_{2\theta_{23}^{}}
\nonumber\\
&& + \frac{\alpha^2 \widehat A ( \eta -c^2_{\theta_{12}^{}}
)^2}{ \widehat C^5 ( 1 + \widehat A + \widehat C)} \Bigm\{
-( 1 + \widehat A )^2 \Bigr[ 1 + 2 \widehat A + \widehat A^2
(3 - \widehat A)\Bigr] + \widehat C + \widehat A \widehat C (
3 + \widehat A + \widehat A^3 )  \nonumber\\
&& -2 \widehat A^2 c^4_{\theta_{13}^{}} (13 -3 \widehat A
- 3 \widehat C ) -\widehat A c^2_{\theta_{13}^{}} \Bigr[-9
+7 \widehat C -24 \widehat A +\widehat A \widehat C + \widehat A^2 ( -11 + 4 \widehat A \nonumber \\ &&
+ 4 \widehat C)\Bigr]
\Bigm\}
c^4_{\theta_{13}^{}} s^2_{2 \theta_{23}^{}}
\biggr\}\frac{\sin^2
\widetilde F_-^{}}{\epsilon^2}
+\frac{8 \alpha {\cal J}}{\epsilon \widehat C (1 + \widehat A +\widehat C)}
(\cos \widetilde F^{}_{+} - \cos \widetilde F^{}_{-}
) \sin \widetilde F^{}_{-}\;,
\label{eq:Ptm}
\end{eqnarray}
where the expressions of $\widetilde F^{}_{\pm}$ and $T$ are shown
in Eqs.~(\ref{eq:Fpm}) and (\ref{eq:T}), respectively.
Taking $\eta= \cos^2 \theta^{}_{12}$ in
Eq.~(\ref{eq:Ptm}), the form of
$\widetilde P_{\tau\mu}^{}$ reduces to
\begin{eqnarray}
\widetilde P^{}_{\tau \mu}  & \approx &
 [\frac{  (1 - \widehat A + \widehat C  ) ^2}
{2 \widehat C^2  (1 + \widehat A + \widehat C  )^2} c^{4}_{\theta_{13}^{}}
s^{2}_{2\theta_{23}^{}}
+ \frac{16 \alpha \widehat A  (1 - \widehat A + \widehat C  )
}{\widehat C^2  (1 + \widehat A + \widehat C  )^3} c^{2}_{\theta_{13}^{}}
 c^{}_{2 \theta_{23}^{}} {\cal J} \cot\delta
\Bigr ] (1 - \cos\widetilde F_{+}^{} \cos\widetilde F_{-}^{}
 )\nonumber \\
&& -  \biggm\{ \frac{  (1 - \widehat A + \widehat C  )
}{8 \epsilon \widehat C^2  (1 + \widehat A + \widehat C  )}
 (1 + \widehat A - \widehat C -2 \alpha c^{}_{2\theta_{12}^{}} )
 (c^{}_{2\theta_{13}^{}} - \widehat A - 3 \widehat C  )
c^{2}_{\theta_{13}^{}} s^{2}_{2\theta_{23}^{}} \nonumber \\
&&  + 2 \alpha c^{}_{2\theta_{23}^{}} {\cal J}\cot\delta
\Bigr [\frac{1 + \widehat A + \widehat C}{\epsilon \widehat C^2} -
\frac{8 \widehat A (\widehat A c^{2}_{\theta_{13}^{}} + \widehat C
+ \widehat A \widehat C)}{\epsilon \widehat C^2
 (1 + \widehat A + \widehat C  )^2} \Bigr]  \biggm\}
\sin\widetilde F_{+}^{} \sin\widetilde F_{-}^{} \nonumber \\
&& +\biggm\{-\frac{\widehat A  (1 - \widehat A - \widehat C)}{\widehat C  (1 +
\widehat A + \widehat C  )} c^{4}_{\theta_{13}^{}} s^{2}_{2\theta_{23}^{}}
+ \frac{\alpha  (1 - \widehat A - \widehat C  )}{\widehat C}
c^{}_{2\theta_{12}^{}} c^{2}_{\theta_{13}^{}}
s^{2}_{2\theta_{23}^{}}
+ \frac{\alpha^2  (1 + \widehat A)}{\widehat C} s^{2}_{2 \theta_{12}
^{}} s^{2}_{\theta_{13}^{}}
\nonumber \\
&& + s^{2}_{\theta_{13}^{}} s^{2}_{2 \theta_{23}^{}}
\frac{\alpha^2  (1 + \widehat A + \widehat C  )}{\widehat C  (1 - \widehat A + \widehat C
 )}
- \alpha^2 s^{2}_{2\theta_{12}^{}} s^{2}_{2\theta_{23}^{}}T
+ \frac{4 \alpha  (1 - \widehat A - \widehat C- 2\widehat A^2 - 2\widehat A \widehat C
 )}{\widehat C  (1 + \widehat A + \widehat C  )}
{\cal J} c^{}_{2\theta_{23}^{}}\cot\delta
\nonumber \\
&&  + 4 \alpha^2 c^{}_{2\theta_{12}^{}}  c^{}
_{2\theta_{23}^{}} {\cal J} \cot\delta
 \Bigr[\frac{4}{ c^{2}_{\theta_{13}^{}}
 (1 - \widehat A + \widehat C  )}
-\frac{ (1 - \widehat A - \widehat C  )^2
 (1 + 2 c^{2}_{\theta_{13}^{}} + \widehat A + 3 \widehat C)
}{2 \widehat C s^{2}_{\theta_{13}^{}}  (1 + \widehat A + \widehat C  )
^2 } \nonumber\\
&& - \frac{2
}{\widehat C  (1 + \widehat A + \widehat C  )}  \Bigr]
-\frac{\alpha^2  (1 + \widehat A  )}{2 \widehat C}
s^{2}_{2\theta_{12}^{}} s^{2}_{\theta_{13}^{}}
s^{2}_{2\theta_{23}^{}} c^{}_{2\delta}
\biggm\}\frac{\sin^2 \widetilde F_{-}^{}} {\epsilon^2}
\nonumber\\ &&- \frac{8 \alpha {\cal J} }{\epsilon \widehat C
 (1 + \widehat A + \widehat C
 )}  (\cos \widetilde F^{}_{-} - \cos \widetilde F^{}_{+}
 ) \sin \widetilde F^{}_{-}\;.
\label{eq:Ptm_new}
\end{eqnarray}
The aboslute error of the above $\widetilde P_{\tau\mu}^{}$ for a wide range of neutrino energies and baseline lengths has been shown in Fig.~\ref{fig:2D}.

\section{Mappings of $\widetilde{\theta}^{}_{ij}$ and $\widetilde{J}$ for a generic $\eta$}
Now we show the mapping of three mixing angles
and the Jarlskog invariant with an arbitrary $\eta$.
Comparing between Eq.~(\ref{eq:Pab_dis}) with
$\alpha=\beta=e$ and Eq.~(\ref{eq:Pee}), or similarly between Eq.~(\ref{eq:Pab_app}) with $(\alpha, \beta) = (\mu, e)$ and Eq.~(\ref{eq:Pme}), one can obtain relations similar to Eq.~(\ref{eq:map_ee}) but for
an arbitrary $\eta$. Then, based on the expressions for
$\xi_i^{ee}$ and $\xi_i^{\mu e}$ in Appendix A,
we can get
\begin{eqnarray}
s^{2}_{\widetilde\theta_{13}^{}}   & \approx &
\frac{
(1 + \widehat A + \widehat C )}{\widehat C (1 - \widehat A + \widehat C
)}s^{2}_{\theta_{13}^{}} + \frac{\alpha\widehat A (\eta - c^{2}_{\theta_{12}^{}})}{2 \widehat C^3}
s^{2}_{2 \theta_{13}^{}}
+ \frac{\alpha^2 \widehat A (\eta - c^{2}_{\theta_{12}^{}})^2
(2 -\widehat A c^{}_{2 \theta_{13}^{}} - \widehat A^2)
}{4 \widehat C^5} s^{2}_{2 \theta_{13}^{}}
\nonumber\\&&
-  \frac{\alpha^2 (1 - \widehat A - \widehat C )
(1 - \widehat A^2 + 3 \widehat C - \widehat A \widehat C )
}{4 \widehat C^3 (1 + \widehat A + \widehat C )^2} s^{2}_{2 \theta_{12}^{}}
c^{2}_{\theta_{13}^{}} \;,
\label{eq:map_12_eta}
\end{eqnarray}
\begin{eqnarray}
 s^{2}_{\widetilde\theta_{12}^{}}  & \approx &
 \frac{1 + 2\epsilon  + \widehat A - \widehat C
- 2 \alpha c^{}_{2\theta_{12}^{}} }{4 \epsilon} +
\frac{\alpha (\eta - c^{2}_{\theta_{12}^{}}) (1 + \widehat A -\widehat C)
(1 - \widehat A -\widehat C)}{8 \epsilon \widehat C}
\nonumber\\   && - \frac{\alpha^2
\widehat A (2 + 3 \widehat A - 6 c^{2}_{\theta_{13}^{}} \widehat A
+ \widehat A^2 + 6 \widehat C - \widehat A \widehat C )}
{2 \epsilon \widehat C(1 - \widehat A + \widehat C )^2 (1 + \widehat A + \widehat C
)} s^{2}_{2\theta_{12}^{}} s^{2}_{\theta_{13}^{}} -\frac{
\alpha^2 \widehat A^2 (\eta - c^{2}_{\theta_{12}^{}})^2
}{8 \epsilon \widehat C^3}  s^{2}_{2 \theta_{13}^{}} \;,
\label{eq:map_13_eta}
\end{eqnarray}
\begin{eqnarray}
 s^{2}_{\widetilde\theta_{23}^{}}   & \approx &
 s^{2}_{\theta_{23}^{}} - \frac{8 \alpha
 (1 - \widehat A - \widehat C )
(1 + \widehat A + \widehat C + 2\alpha c^{}_{2\theta_{12}^{}}
)} {s^{2}_{2 \theta_{13}^{}}
(1 + \widehat A + \widehat C )^2} {\cal J} \cot\delta+ \frac{\alpha^2
(1 - \widehat A - \widehat C )^2}{
4s^{2}_{\theta_{13}^{}}
(1 + \widehat A + \widehat C )^2} s^{2}_{2\theta_{12}^{}}
c^{}_{2\theta_{23}^{}}
\nonumber\\ && -\frac{ 8 \alpha^2  (\eta - c^{2}
_{\theta_{12}^{}}) (1 + \widehat C) (1 - \widehat A - \widehat C)
}{s^{2}_{2 \theta_{13}^{}} \widehat C (1 +
\widehat A + \widehat C)} {\cal J} \cot\delta \;,
\label{eq:map_23_eta}
\end{eqnarray}
\begin{eqnarray}
\widetilde {\cal J}   & \approx &
 \frac{2 \alpha {\cal J}}{\epsilon \widehat C
(1 + \widehat A + \widehat C )} + \frac{2 \alpha^2
(1 - \widehat A - \widehat C ) }
{\epsilon \widehat C (1 + \widehat A + \widehat C )^2} {\cal J}c^{}
_{2\theta_{12}^{}} \nonumber\\ &&
-\frac{\alpha^2 (\eta - c^{2}_{\theta_{12}^{}}) {\cal J} }{
\epsilon \widehat C^3 (1 + \widehat A + \widehat C) } (1 - 4 \widehat A
c^{}_{2 \theta_{13}^{}} + 3 \widehat A^2 - \widehat C + \widehat A \widehat C) \;,
\label{eq:map_J_eta}
\end{eqnarray}
which reduce to the results given in Eqs.~(\ref{eq:map_mix}) and (\ref{eq:map_J}) if $\eta=\cos^2 \theta^{}_{12}$ is taken.
\newpage
\section*{Acknowledgements}

The authors would like to thank Professor Zhi-zhong Xing for helpful discussions and partial involvement at the early stage of this work, which was supported in part by the National Natural Science Foundation of China under Grant Nos. 11135009 and 11305193, by the Strategic Priority Research Program of the Chinese Academy of Sciences under Grant No. XDA10010100, by the National Recruitment Program for Young Professionals and the CAS Center for Excellence in Particle Physics (CCEPP).

\end{document}